\documentclass[fleqn,reqno]{amsart}

\usepackage{graphicx,amscd,amsmath,amssymb,verbatim}
\usepackage[dvips]{hyperref}
\usepackage[TS1,OT1,T1]{fontenc}

\begin{document}

\title[Reciprocal relativity of noninertial frames]{Reciprocal relativity of noninertial frames and the quaplectic group}
\author{Stephen G. Low}
\address{hp}
\email{Stephen.Low@hp.com}
\date{\today}
\begin{abstract}
Newtonian mechanics has the concept of an absolute inertial rest
frame. Special relativity eliminates the absolute rest frame but
continues to require the absolute inertial frame. General relativity
solves this for gravity by requiring particles to have locally inertial
frames on a curved position-time manifold. The problem of the absolute
inertial frame for other forces remains.\ \ We look again at the
transformations of frames on an extended phase space with position,
time, energy and momentum degrees of freedom.\ \ Under nonrelativistic
assumptions, there is an invariant symplectic metric and a line
element $d t^{2}$.\ \ Under special relativistic assumptions the
symplectic metric continues to be invariant but the line elements
are now $-d t^{2}+\frac{1}{c^{2}} d q^{2}$ and $d p^{2}-\frac{1}{c^{2}}d
e^{2}$.\ \ Max Born conjectured that the line element should be
generalized to the pseudo- orthogonal metric $-d t^{2}+\frac{1}{c^{2}}
d q^{2}+\frac{1}{b^{2}}(d p^{2}-\frac{1}{c^{2}}d e^{2})$.\ \ The
group leaving these two metrics invariant is the pseudo-unitary
group of transformations between noninertial frames. We show that
these transformations eliminate the need for an absolute inertial
frame by making forces relative and bounded by $b$ and so embodies
a relativity that is {\itshape reciprocal} in the sense of Born.
The inhomogeneous version of this group is naturally the semidirect
product of the pseudo-unitary group with the nonabelian Heisenberg
group.\ \ This is the quaplectic group. The Heisenberg group itself
is the semidirect product of two translation groups.\ \ This provides
the noncommutative properties of position and momentum and also
time and energy that are required for the quantum mechanics that
results from considering the unitary representations of the quaplectic
group.\ \ 
\end{abstract}
\maketitle

\section{Introduction}

This year is the one hundredth anniversary of Einstein's special
relativity theory that changed our concept of space and time by
combining the separate concepts of space and time into a unified
four dimensional space-time. Newtonian mechanics has the concept
of an absolute inertial rest frame that all observers agree upon,
and furthermore all observers agree upon a single universal concept
of time.\ \ Special relativity eliminates the absolute rest frame
and the universal concept of time, but an absolute inertial frame
that all observers agree upon continues to exist. 

General relativity eliminates the need for an absolute inertial
frame for gravitational forces by requiring particles under the
influence of gravity to be in a curved space-time manifold. In this
curved space-time, gravitating particles follow geodesics and therefore
are locally inertial.\ \ In this sense, gravity is no longer a {\itshape
force} that causes particles to transition to noninertial frames.
This sidesteps the issue of a global inertial frame by eliminating
the concept of noninertial frames as all frames for particles that
are only under the influence of gravity are locally inertial. 

However, if other forces are considered, the question of the absolute
inertial frame reemerges. Return again to the case where the underlying
space-time manifold is flat. Special relativity eliminates the problem
of the absolute rest frame, simply by showing that there is no absolute
rest frame. Velocities are meaningful only between observers associated
with particle states and the relative velocity is bounded by $c$.\ \ Can
we follow this approach of special relativity and simply eliminate
the concept of a global inertial frame?\ \ The answer to this, remarkably,
is yes.\ \ Forces can be made relativistic also so that rates of
change of momentum are meaningful only between particle states (and
not relative to some absolute inertial frame) and furthermore are
bounded by a universal constant $b$.\ \ However, with the elimination
of the absolute inertial frame, space-time takes on even more unusual
properties than the four dimensional continuum introduced by special
relativity.

The first step is to formulate the transformations of noninertial
frames as the action of a more general group. The key to this is
the observation that nonrelativistic Hamilton's mechanics is a continuous
unfolding of a canonical transformation. As this represents all
possible motions of a classical point particle, these frames are
generally noninertial.\ \ Hamilton's equations may be formulated
as a group of transformations of frames on time-position-momentum-energy
space. These equations leave invariant the symplectic metric, and
are therefore canonical, as well as the invariant nonrelativistic
line element, $d s^{2 }=d t^{2}$.\ \ These transformations describe
general noninertial frames in the nonrelativistic context where
there is an absolute inertial rest frame. 

Next, consider frames on time-position-momentum-energy space under
special relativistic transformations. Special relativity\ \ eliminates
the concept of the absolute rest frame by requiring the invariance
of the line element\ \ ${\mathit{ds}}^{2}=-{\mathit{dt}}^{2}+\frac{1}{c^{2}}{\mathit{dq}}^{2}$
on position-time space and the line element ${\mathit{d}\mu }^{2}=-\frac{1}{c^{2}}{\mathit{de}}^{2}+{\mathit{dp}}^{2}$
on energy-momentum space.\ \ However, requiring these line elements
to be independently invariant requires the canonical frame to be
an inertial frame. To regain the general noninertial transformation
equations of the nonrelativistic case, we must combine these two
line elements into a single line element defined by the Born-Green
metric \cite{born-einstein}\ \ ${\mathit{ds}}^{2}=-{\mathit{dt}}^{2}+\frac{1}{c^{2}}{\mathit{dq}}^{2}+\frac{1}{b^{2}}({\mathit{dp}}^{2}-\frac{1}{c^{2}}{\mathit{de}}^{2})$.\ \ Now
we have two metrics that must be invariant under the transformations,\ \ the
symplectic metric and the Born-Green orthogonal metric.\ \ The required
group in $n$ dimensions is $\mathcal{U}( 1,n) $ with $n=3$ corresponding
to the usual physical case.

The introduction of the Born-Green metric enables the group of transformations
to again include noninertial frames on time-position-momentum-energy
space.\ \ The full group, $\mathcal{U}( 1,n) , $describes the transformations
between frames that are canonical and noninertial. The remarkable
fact is that, for these transformations,\ \ {\itshape there is no
longer an absolute inertial frame nor an absolute rest frame}. Forces
have become relative and are bounded by $b$.\ \ Velocities continue
to be relative and are bounded by $c$. We call this reciprocal relativity.\ \ There
is also no longer an absolute concept of position-time space that
all observers agree on.\ \ The determination of the position-time
subspace of the time-position-momentum energy space has become observer
frame dependent. Under extreme noninertial conditions, that is,
very strongly interacting particles, all of the time, position,
momentum and energy\ \ degrees of freedom are mixed under this group
of transformations that takes us from one noninertial observer to
another.\ \ Simply put, momentum and energy can be transformed into
position and time.$\text{}$

This generalization is very analogous to time becoming relative
in the special relativity theory. In Newtonian mechanics, the Lorentz
group contracts in the limit $c\rightarrow \infty $ to the Euclidean
group. The Euclidean group acting on the frame leaves time invariant.
Consequently, time is absolute in the sense that the determination
of the time subspace of the position-time manifold is common to
all observers. Under the action of the Lorentz group in special
relativity,\ \ time is observer frame dependant and so there is
no longer an absolute concept of time.

These transformations reduce to the canonical transformations between
inertial frames of special relativity if the rate of change of momentum
between the observers is zero. In this case, the transformation
equations reduce to the usual special relativity Lorentz group transformations
that act separately on position-time\ \ and energy-momentum space.
The concept of a position-time space and energy-momentum space,
on which all inertial observers agree, reemerges. 

Up until this point we have been discussing the homogeneous group
that generalizes the concept of the Lorentz group to noninertial
frames. The basic wave or field equations of special relativistic
quantum mechanics arise from the study of the unitary irreducible
representations of the Poincar\'e group \cite{wigner1}.\ \ The Poincar\'e
group is the semidirect product of the Lorentz group with the abelian
translation group.\ \ The Hermitian representations of the eigenvalue
equations of the Casimir invariant operators define the\ \ free
particle (or inertial frame) basic wave equations, the Klein-Gordon,
Dirac, Maxwell equations and so forth.\ \ \ 

Simply considering the semidirect product of\ \ the $\mathcal{U}(
1,n) $ group of reciprocal relativity\ \ with the abelian translation
group does not work. First, a basic principle of quantum mechanics
is that the position and momentum and energy and time degrees of
freedom do not commute. In fact, in nonrelativistic quantum mechanics,
position and momentum are realized as the Hermitian representation
of the algebra associated with the\ \ unitary irreducible representations
of the Heisenberg group.\ \ The Heisenberg group itself is a real
matrix Lie group that is the semidirect product of two translation
groups, $\mathcal{H}( n) =\mathcal{T}( n) \otimes _{s}\mathcal{T}(
n+1) $.\ \ As a real matrix group, the Heisenberg Lie algebra is
\[
\left[ P,Q\right] =I,\ \ \ \left[ E,T\right] =-I
\]

The quaplectic group is then defined as the semidirect product
\[
\mathcal{Q}( 1,n) =\mathcal{U}( 1,n) \otimes _{s}\mathcal{H}\left(
n+1\right) .
\]

\noindent It is important to emphasize that the non-commutative
property of the position and momentum and the energy and time degrees
of freedom therefore appear already in the classical (i.e. non quantum)
formulation. That is, this may be viewed simply as a noncommutative
geometry on the $2n+4$ dimensional space $\mathcal{Q}( 1,n) /\mathcal{S}\mathcal{U}(
1,n) $ analogous to Minkowski space $\mathcal{P}( 1,n) /\mathcal{S}\mathcal{O}(
1,n) $. For $n=3$, this space has 10 dimensions. 

The quantum theory of the particles in the noninertial frames then
follows directly by considering the unitary irreducible representations
of the quaplectic group.\ \ Note that the expected Heisenberg relations
follow directly from the Hermitian representations of the algebra
corresponding to unitary irreducible representations of the Heisenberg
group. The\ \ wave equations that result are derived in\ \ \cite{low-Casimir
of Quaplectic}.\ \ By way of example, the equivalent in this theory
of the scalar equation is the relativistic oscillator.\ \ It is
derived directly from the Hermitian representations of the eigenvalue
equation for the Casimir invariants in the scalar representation.
The Casimir invariant is by definition an invariant of the quaplectic
group, and in particular, is invariant under Heisenberg nonabelian
{\itshape translations}. This is exactly the method used to compute
the free particle wave equations from the Poincar\'e group. 

Returning again to the classical (ie non-quantum), but nonabelian
theory, there is another fundamental motivation for why the inhomogeneous
group cannot be the abelian translations. Clearly, if a general
relativity type theory is to be considered in this framework, it
must be possible to admit manifolds that are not flat. Schuller
\cite{Schuller} has proven a {\itshape no-go} theorem that states
that a manifold that is a tangent bundle with a symplectic and (Born-Green)
orthogonal metric must be flat.\ \ One way to invalidate the no-go
theorem is to make the manifold noncommutative.\ \ For the physical
case $n=3$, this results in a 10 dimensional noncommutative curved
manifold. How to formulate this differential geometry is an open
question and the topic of a subsequent investigation.\ \ 

A final observation about the choice of the Heisenberg group as
the normal subgroup of the semidirect product.\ \ Unlike the translation
group where we can construct the inhomogeneous general linear group,
the group "$\mathcal{G}\mathcal{L}( n) \otimes _{s}\mathcal{H}(
n) $"\ \ does not exist. This is because the automorphisms of the
quaplectic group require the homogeneous group in this semidirect
product to be the symplectic group, or one of its subgroups. Thus
the requirement for the normal subgroup to be the Heisenberg group
automatically requires the symplectic metric that has been fundamental
in our discussion.

This paper seeks to motivate the above description in the simplest
possible manner. It looks at a system that has only one position
dimension and traces the above arguments through to show how the
quaplectic group arises.\ \ The companion paper \cite{low-Casimir
of Quaplectic} then determines the unitary representations of the
$n$ dimensional quaplectic group, the associated Hilbert spaces
and the field equations that arise from the Hermitian representation
of the eigenvalue equations of the Casimir invariant operators.

\section{Homogeneous group: Relativity}

\subsection{Basic special relativity}\label{Section Basic special
relativity}

The special relativity transformation equations between inertial
observers. As we are considering the one dimensional case\ \ \ $x=\{t,q\}\in
\mathbb{M}\simeq \mathbb{R}^{2}$.\ \ The global transforms are
\[
\tilde{ t}={\left( 1-{\left( \frac{v}{c}\right) }^{2}\right) }^{-1/2}\left(
t+\frac{v}{c^{2}} q\right) ,\ \ \tilde{ q}={\left( 1-{\left( \frac{v}{c}\right)
}^{2}\right) }^{-1/2}\left(  q+v t\right) 
\]

\noindent These expressions are made local by lifting to the cotangent
space acting on a frame $\{d t, d q\}$.\ \ \ 
\begin{equation}
d\tilde{ t}= {\left( 1-{\left( \frac{v}{c}\right) }^{2}\right) }^{-1/2}\left(
d t+\frac{v}{c^{2}}d q\right) , d\tilde{ q}={\left( 1-{\left( \frac{v}{c}\right)
}^{2}\right) }^{-1/2}\left(  d q+v d t\right) %
\label{basic special relativity local transformations}
\end{equation}

\noindent This may be written in matrix notation as $d\tilde{x}
= \Lambda ( v)  d x$ where $d x=\{d t,d q\}$, $d x\in {T^{*}}_{x}\mathbb{M}$
and\ \ $\Lambda ( v) $ is a $2\times 2$ matrix 
\begin{equation}
 \Lambda ( v) ={\left( 1-{\left( \frac{v}{c} \right) }^{2}\right)
}^{-1/2}\left( \begin{array}{ll}
 1 & \frac{v}{c^{2}} \\
 v & 1
\end{array}\right) %
\label{special relativity lambda matrix}
\end{equation}

\noindent These transformations leave invariant the line elements
${\mathit{ds}}^{2}=-{\mathit{dt}}^{2}+\frac{1}{c^{2}}{\mathit{dq}}^{2}$.
The usual addition law for special relativity results directly from
the group composition law
\begin{equation}
\Lambda ( \tilde{\tilde{v}}) =\Lambda ( \tilde{v}) \cdot \Lambda
( v) ,\ \ \ \ \Lambda ^{-1}( v) = \Lambda ( -v) %
\label{special relativity group transformation law}
\end{equation}

\noindent Multiplying out the matrices gives the expected result
\begin{equation}
\tilde{\tilde{v}}=\frac{\tilde{v}+v}{1+\tilde{v}v/c^{2}}%
\label{special relativity velocity addition}
\end{equation}

\noindent \noindent Physically, if $v$ is the velocity between inertial
frame of observers 1 and 2, $\tilde{v}$ the relative velocity between
inertial frames of observers 2 and 3, then $\tilde{\tilde{v}}$ given
by this expression is the relative velocity observed between inertial
frames of observers 1 and 3.\ \ 

\noindent \noindent Alternatively, from (1\noindent ), 
\begin{equation}
\frac{d\tilde{ q}}{d\tilde{ t}}= \left(  d q+v d t\right) /\left(
d t+\frac{v}{c^{2}}d q\right) = \left(  \frac{d q}{d t}+v\right)
/\left( 1+\frac{v}{c^{2}}\frac{d q}{d t}\right)  %
\label{basic special relative rate equation}
\end{equation}

\noindent Comparing this equation with (4) leads to the identification
of $v$ with the rate of change of position, $d q/d t$.

Now the nonrelativistic case is given simply by requiring the limit\ \ 
\begin{equation}
\Phi ( v) =\operatorname*{\lim }\limits_{c\,\rightarrow \:\infty
}\Lambda ( v)  =\left( \begin{array}{ll}
 1 & 0 \\
 v & 1
\end{array}\right) %
\label{phi of v matrix}
\end{equation}

\noindent Multiplying out the matrices 
\begin{equation}
\Phi ( \tilde{\tilde{v}}) =\Phi ( \tilde{v}) \cdot \Phi ( v) , \Phi
^{-1}( v) = \Phi ( -v) %
\label{Newtonian group transformation law}
\end{equation}

\noindent gives the expected result 
\begin{equation}
\tilde{\tilde{v}}=\tilde{v}+v%
\label{newtonian velocity addition}
\end{equation}

\noindent and so the nonrelativistic transformations between inertial
frames are\ \ $d\tilde{ x}=\Phi ( v) d x$\ \ is just the expected
\begin{equation}
\begin{array}{l}
 d\tilde{t}=d t \\
 d\tilde{q}=d q + v d t
\end{array}%
\label{newtonian nonrelativistic inertial}
\end{equation}

\noindent These equations leave invariant the classical line element
\begin{equation}
d s^{2}= \operatorname*{\lim }\limits_{c\,\rightarrow \:\infty }\left(
-{\mathit{dt}}^{2}+\frac{1}{c^{2}}{\mathit{dq}}^{2}\right) =-d t^{2}%
\label{Newtonian line element}
\end{equation}

\noindent For this reason, we say that Newtonian mechanics has\ \ a
concept of absolute time.\ \ Then, as in the special relativistic
case (5), divide by $d t$\ \ to obtain the usual velocity addition
(8)
\begin{equation}
\frac{d\tilde{ q}}{d\tilde{ t}}= \left(  \frac{d q}{d t}+v\right)
\label{non-relativistic inertial rate transformation}
\end{equation}

The key difference between special relativity and Newtonian mechanics
is that the latter has a universal `inertial rest frame' whereas
in the former, velocity is relative and so there is no absolute
rest frame. However, in special relativity, there continues to be
a global inertial frame.\ \ 

\subsection{Inertial canonical frame transformations of special
relativity}

The above basic arguments may be repeated on time-position-momentum-energy
space. In the following argument, this combination is somewhat formal
as these transformations never mix. However, it motivates the section
that follows. 

The special relativity transformation equations between inertial
observers and consider also the momentum, energy equations. $z=\{t,q,p,e\}\in
\mathbb{P}\simeq \mathbb{R}^{4}$. 
\begin{gather*}
\tilde{ t}={\left( 1-{\left( \frac{v}{c}\right) }^{2}\right) }^{-1/2}\left(
t+\frac{v}{c^{2}} q\right) ,\ \ \tilde{ q}={\left( 1-{\left( \frac{v}{c}\right)
}^{2}\right) }^{-1/2}\left(  q+v t\right) 
\end{gather*}
\[
\tilde{ p}={\left( 1-{\left( \frac{v}{c}\right) }^{2}\right) }^{-1/2}\left(
p+\frac{v}{c^{2}} e\right) , \tilde{e}={\left( 1-{\left( \frac{v}{c}\right)
}^{2}\right) }^{-1/2}\left( e+v p\right) 
\]

In these expressions, $t$ is time, $q$ is position, $p$ is momentum
and, with a little unconventional notation, $e$ is energy.\ \ Again,
we make the expressions local, applying the same argument also to
the momentum and energy degrees of freedom, by lifting to the cotangent
space with a frame $\{d t, d q,d p,d e\}$.\ \ \ 
\begin{gather*}
d\tilde{ t}={\left( 1-{\left( \frac{v}{c}\right) }^{2}\right) }^{-1/2}\left(
d t+\frac{v}{c^{2}}d q\right) , d\tilde{ q}={\left( 1-{\left( \frac{v}{c}\right)
}^{2}\right) }^{-1/2}\left(  d q+v d t\right) 
\end{gather*}
\[
d\tilde{ p}={\left( 1-{\left( \frac{v}{c}\right) }^{2}\right) }^{-1/2}\left(
d p+\frac{v}{c^{2}} d e\right) ,\ \ d\tilde{e}={\left( 1-{\left(
\frac{v}{c}\right) }^{2}\right) }^{-1/2}\left( d e+v d p\right)
\]

\noindent This may be written in matrix notation as $d\tilde{z}
= \Gamma ( v)  d z$ where $d z=\{d t,d q,d p, d e\}$, $d z\in {T^{*}}_{z}\mathbb{P}$
and\ \ $\Gamma ( v) $ is a $4\times 4$ matrix 
\[
\Gamma ( v) =\left( \begin{array}{ll}
 \Lambda ( v)  & 0 \\
 0 & \Lambda ( v) 
\end{array}\right) ,\ \ \ \ \ \Lambda ( z) ={\left( 1-{\left( \frac{v}{c}\right)
}^{2}\right) }^{-1/2}\left( \begin{array}{ll}
 1 & \frac{v}{c^{2}} \\
 v & 1
\end{array}\right) 
\]

\noindent Expanding this out gives the $4\times 4$ matrix
\begin{equation}
\Gamma ( v) ={\left( 1-{\left( \frac{v}{c}\right) }^{2}\right) }^{-1/2}\left(
\begin{array}{llll}
 1  & \frac{v}{c^{2}} & 0 & 0 \\
  v  & 1  & 0 & 0 \\
 0 & 0 & 1  & \frac{v}{c^{2}} \\
 0 & 0 & v  & 1 
\end{array}\right) 
\end{equation}

As the matrix is simply the direct sum, the subspaces spanned by
$\{d t,d q\}$ and $\{d p, d e\}$ do not mix.\ \ These transformations
leave invariant the line elements ${\mathit{ds}}^{2}=-{\mathit{dt}}^{2}+\frac{1}{c^{2}}{\mathit{dq}}^{2}$
and independently ${\mathit{d}\mu }^{2}=-\frac{1}{c^{2}}{\mathit{de}}^{2}+{\mathit{dp}}^{2}$.
The symplectic metric continues to be invariant also
\begin{equation}
\begin{array}{rl}
 {}^{t} d z \cdot \zeta \cdot  d z & =-d e \wedge d t+d p\wedge
d q
\end{array}%
\label{symplectic metric defn}
\end{equation}

\noindent Thus, in addition to being inertial, the frames are canonical
in the sense that the symplectic metric has the simple form given.\ \ The
appearance of the symplectic metric is not arbitrary but directly
required if a semidirect product group with the Heisenberg group
as the normal subgroup is to be constructed for the inhomogeneous
case. This is described in Section 3.3. 

The usual addition law for special relativity results directly from
the group composition law, $\Lambda ( \tilde{\tilde{v}}) =\Lambda
( \tilde{v}) \Lambda ( v) $.\ \ Multiplying out the matrices gives
the expected result (4).\ \ Physically, if $v$ is the velocity between
inertial frame of observers 1 and 2, $\tilde{v}$ the relative velocity
between inertial frames of observers 2 and 3, then $\tilde{\tilde{v}}$
given by this expression is the relative velocity observed between
inertial frames of observers 1 and 3.\ \ The identification with
velocity is verified by
\begin{gather}
\frac{d\tilde{ q}}{d\tilde{ t}}=\ \ \left(  \frac{d q}{d t}+v\right)
/\left( 1+\frac{v}{c^{2}}\frac{d q}{d t}\right)  
\end{gather}
\begin{equation}
\frac{\partial \tilde{ e}}{\partial \tilde{ p}}=\left( \frac{\partial
e}{\partial  p}+v \right) /\left( 1+\frac{v}{c^{2}}  \frac{\partial
e}{\partial  p}\right) %
\label{special relativity phase space velocity}
\end{equation}

\noindent Now the nonrelativistic case is given simply by taking
the limit\ \ 
\begin{equation}
\Phi ( v) =\operatorname*{\lim }\limits_{c\,\rightarrow \:\infty
}\Gamma ( v)  =\left( \begin{array}{llll}
 1  & 0 & 0 & 0 \\
  v  & 1  & 0 & 0 \\
 0 & 0 & 1  & 0 \\
 0 & 0 & v  & 1 
\end{array}\right) 
\end{equation}

\noindent and\ \ the nonrelativistic transformations between inertial
frames,\ \ $d\tilde{ z}=\Phi ( v) d z$ are the expected
\begin{equation}
\begin{array}{l}
 d\tilde{t}=d t \\
 d\tilde{q}=d q + v d t \\
 d\tilde{p}=d p  \\
 d \tilde{e} = d e + v d p 
\end{array}%
\label{nonrelativistic inertial}
\end{equation}

\noindent These equations leave invariant the line element $d s^{2}=-d
t^{2}$ (10) and also the symplectic metric (13).\ \ $\Phi ( v) $
is just a\ \ realization of the one\ \ dimensional translation group
and so again, the velocity addition law is\ \ $\Phi ( \tilde{\tilde{v}})
=\Phi ( \tilde{v}) \Phi ( v) $ and\ \ it follows that $\tilde{\tilde{v}}=\tilde{v}+v$
. 

Again, these transformation equations are for frames that are both
inertial and canonical. For heuristic purposes, note that (15) leads
to the identification $v=\frac{d q}{d t}=\frac{\partial e}{\partial
p} $. We will return to this more carefully shortly.\ \ 

\subsection{Noninertial frame transformations of nonrelativistic
mechanics}

Nonrelativistic Hamilton's mechanics does not have the requirement
for frames to be inertial. Given the Hamiltonian function, particles
with complex noninertial motion can be described. All point particle
states\ \ in classical mechanics, and the associated noninertial
frames, must comply with Hamilton's equations.\ \ The basic idea
of Hamilton's mechanics is that the particle motion is the continuous
unfolding of a canonical transformation. Hamilton's equations are
\begin{equation}
v=\frac{d q( t) }{d t}=\frac{\partial H( p,q,t) }{\partial p}, f=\frac{d
p( t) }{d t}=-\frac{\partial H( p,q,t) }{\partial q}, r=\frac{\partial
H( p,q,t) }{\partial t}%
\label{Hamiltons equations}
\end{equation}

\noindent Consider the co-ordinate transformation $\tilde{z}=\varphi
( z) $. Then, with simple matrix notation
\begin{equation}
d\tilde{z}=\frac{\partial \varphi ( z) }{\partial z}\mathit{dz}
= \Phi ( v,f,r) d z%
\label{general transformation}
\end{equation}

\noindent and so in component form 
\begin{equation}
\frac{\partial \varphi ^{\alpha }}{\partial z^{\beta }}\ \ ={\Phi
( v,f,r) }_{\beta }^{\alpha },\ \ \ \ \ \ \ \alpha ,\beta =1,2,3,4\text{}.
\end{equation}

\noindent The solution of Hamilton' s equations defines the canonical
transformations $\varphi ^{\alpha }( t,q,p,e) $ that have the form
\begin{equation}
\begin{array}{ll}
 \tilde{t}=\varphi ^{1}( t,q,p,e) =t &   \\
 \tilde{q}=\varphi ^{2}( t,q,p,e) =q+q( t)   & q( 0) =0 \\
 \tilde{p}=\varphi ^{3}( t,q,p,e) =p+p( t)  & p( 0) =0 \\
 \tilde{e}=\varphi ^{4}( t,q,p,e) =e+H( p,q,t)   & H( 0,0,0) =0
\end{array}%
\label{Hamilton transformations}
\end{equation}

\noindent The notation is is being abused slightly using $q,p$ for
the initial points and $q( t) ,p( t) $ for the functions giving
the time evolution. Substituting\ \ (21) into (19) and using Hamilton's
equations (18) gives the result
\begin{equation}
\Phi ( v,f,r) \simeq \left( \begin{array}{llll}
 1 & 0 & 0 & 0 \\
 v & 1 & 0 & 0 \\
 f & 0 & 1 & 0 \\
 r & -f & v & 1
\end{array}\right) %
\label{Hamilton group matrix}
\end{equation}

\noindent Writing out the transformations using\ \ $d\tilde{z}=
\Phi ( v,f,r) d z$ yields
\begin{equation}
\begin{array}{l}
 d\tilde{t}=d t \\
 d\tilde{q}=d q + v d t \\
 d\tilde{p}=d p + f d t  \\
 d \tilde{e} = d e + v d p -f d q + r d t
\end{array}%
\label{nonrelativistic transformation equations}
\end{equation}

Conversely, starting with (23) in (19), one can derive (21) and
Hamilton's equations (18). These equations define the transformations
between noninertial frames. Only noninertial frames that satisfy
these transformations, and hence Hamilton's equations, are physical.

\noindent These transformations leave invariant the line element
$d s^{2}=-d t^{2}= {}^{t} d z\cdot \eta _{c}\cdot d z$ and also
the symplectic metric
\[
\begin{array}{rl}
 {}^{t} d \tilde{z} \cdot \zeta \cdot  d \tilde{z} & =-d\tilde{e}\wedge
\mathit{d}\tilde{t}\mathrm{+}\mathit{d}\tilde{p}\wedge \mathit{d}\tilde{q}
\\
  & =-\left( d e + v d p -f d q + r d t\right) \wedge d t+\left(
d p + f d t \right) \wedge \left( d q + v d t\right)  \\
  & =-d e \wedge d t+d p\wedge d q={}^{t} d z \cdot \zeta \cdot
d z
\end{array}
\]

\noindent In fact, $\Phi $ may be derived from the requirement that
${}^{t}\Phi \cdot \zeta \cdot \Phi  =\zeta $ and ${}^{t}\Phi \cdot
\eta _{c}\cdot \Phi  =\eta _{c}$\ \ where
\begin{equation}
\zeta =\left( \begin{array}{llll}
 0 & 0 & 0 & 1 \\
 0 & 0 & -1 & 0 \\
 0 & 1 & 0 & 0 \\
 -1 & 0 & 0 & 0
\end{array}\right) ,\ \ \ \eta _{c}=\left( \begin{array}{llll}
 -1 & 0 & 0 & 0 \\
 0 & 0 & 0 & 0 \\
 0 & 0 & 0 & 0 \\
 0 & 0 & 0 & 0
\end{array}\right) 
\end{equation}

Certainly the momentum transformation for these frames, that are
canonical but not necessarily inertial, is the form expected.\ \ The
energy transformation has a kinetic term $ v d p$, as in the inertial
equations above, a new work term $-f d q$ and the explicit power
term $r d t$.\ \ 

Direct matrix multiplication defines the group composition laws
and verifies that this is a matrix Lie group
\begin{equation}
\begin{array}{l}
 \Phi ( \tilde{\tilde{v}},\tilde{\tilde{f}},\tilde{\tilde{r}}) =\Phi
( \tilde{v},\tilde{f},\tilde{r}) \cdot \Phi ( v,f,r) =\Phi ( \tilde{v}+v,\tilde{f}+f,\tilde{r}+r+\tilde{v}f-\tilde{f}v)
, \\
 \Phi ^{-1}( v,f,r) =\Phi ( -v,-f,-r) 
\end{array}%
\label{Hamilton group composition law}
\end{equation}

We call this the Hamilton group in one dimension,\ \ $\Phi ( v,f,r)
\in \mathcal{H}a( 1) $.\ \ As in the inertial case, this defines
the generalized addition laws for velocity $v$, force, $f$ and power,
$r$ for three different observers. That is if we have relative $(v,f,r$)
between noninertial observer frames $1$ and $2$\ \ and relative
$(\tilde{v},\tilde{f},\tilde{r})$ between observer frames 2 and
3, then observer frames 1 and 3 are related by $(\tilde{\tilde{v}},\tilde{\tilde{f}},\tilde{\tilde{r}})$
where 
\begin{equation}
\begin{array}{l}
 \tilde{\tilde{v}}=\tilde{v} + v  \\
 \tilde{\tilde{f}}=\tilde{f}+ f\ \  \\
 \tilde{\tilde{r}}= \tilde{r} + v \tilde{f} - f \tilde{v}\ \ +r
\end{array}%
\label{Hamilton rate transformations}
\end{equation}

\noindent  As time is invariant under the transformations, we can
simply divide (23) by $d t$ to obtain.
\begin{equation}
\frac{d\tilde{q}}{d\tilde{t}}=\frac{d q}{d t} + v,\ \ \frac{d\tilde{p}}{d\tilde{t}}=\frac{d
p}{d t} + f,\ \ \frac{d\tilde{e}}{d\tilde{t}}= \frac{d e}{d t} +
v \frac{d p}{d t} - f \frac{d q}{d t} +r%
\label{nonrelativistic rate equation}
\end{equation}

\noindent Comparing with (25), this lead to the identification of
$v$ with rate of change of position, $f$ with rate of change of
momentum and $r$ with rate of change of energy with respect to the
invariant time element $d t$ as asserted.\ \ 

The matrix Lie algebra of $\mathcal{H}a( 1) $ follows directly from
the Lie algebra valued one forms\ \ $d \Phi ( v,f,r) |_{0}=\mathit{dv}
G+d\mathit{f} \mathit{F}\mathit{+}\mathrm{dr} R$ where the matrix
realization of the generators is given by
\begin{equation}
\begin{array}{ll}
 G=\left( \begin{array}{llll}
 0 & 0 & 0 & 0 \\
 1 & 0 & 0 & 0 \\
 0 & 0 & 0 & 0 \\
 0 & 0 & 1 & 0
\end{array}\right) , & F=\left( \begin{array}{llll}
 0 & 0 & 0 & 0 \\
 0 & 0 & 0 & 0 \\
 1 & 0 & 0 & 0 \\
 0 & -1 & 0 & 0
\end{array}\right) , \\
 R=\left( \begin{array}{llll}
 0 & 0 & 0 & 0 \\
 0 & 0 & 0 & 0 \\
 0 & 0 & 0 & 0 \\
 1 & 0 & 0 & 0
\end{array}\right)  &  
\end{array}
\end{equation}

with
\[
G=\frac{\partial \Phi }{\partial v},\ \ F=\frac{\partial \Phi }{\partial
f},\ \ R=\frac{\partial \Phi }{\partial r}
\]

\noindent The Lie bracket of a matrix Lie group is given simply
by $[A,B]=A\cdot B-B\cdot A$ and these may be directly computed
as
\begin{equation}
\left[ G,F\right] =2 R,\ \ \left[ R,F\right] =0,\ \ \left[ R,G\right]
=0%
\label{Hamilton algebra}
\end{equation}

\noindent These may also be viewed as an abstract Lie algebra satisfying
the above commutation relations.\ \ The manner in which this formulation
carries over to Lagrangian mechanics is summarized in Section 5.1.

\subsection{Reciprocal relativity}

This is where we depart from a simple review of standard physics.
We now introduce the Born-Green conjecture of the combined line
element and derive the transformation equations that encompass reciprocal
relativity. 

Two conditions must be satisfied by the transformation equations.\ \ As,
always, the symplectic metric $-\mathit{de}\wedge \mathit{dt}\mathrm{+}\mathit{dp}\wedge
\mathit{dq}$ must be invariant.\ \ \ Furthermore, in the classical
nonrelativistic theory, we have the invariant ${\mathit{dt}}^{2}$
while in the special relativistic\ \ case we have the invariants
$\mathit{-}{\mathit{dt}}^{\mathrm{2}}+\frac{1}{c^{2}}{\mathit{dq}}^{\mathrm{2}}$
and\ \ $-\frac{1}{c^{2}}{\mathit{de}}^{\mathrm{2}}+\text{}{\mathit{dp}}^{\mathrm{2}}$.\ \ Following
Born-Green, this suggests we investigate that case of further combining
these invariants to define a non-degenerate (pseudo) orthogonal
metric on the 4 dimensional space 
\begin{equation}
d s^{2}=-d t^{2}+\frac{1}{c^{2}}d q^{2}+\frac{1}{b^{2}}\left( -\frac{1}{c^{2}}d
e^{2}+d p^{2}\right) = {}^{t}d z\cdot \eta \cdot d z%
\label{quaplectic orthogonal line element}
\end{equation}

\noindent where $\eta =\mathrm{diag}\{-1,1,1,-1\}$.

This is the only new physical assumption that has been introduced
in this paper.\ \ $b$ is a universal constant with the dimensions
of force.\ \ Usually $(c,\hbar ,G)$ are taken to be the independent
dimensional constants. In terms of this dimensional basis, 
\begin{equation}
b=\alpha _{G }\frac{c^{4}}{G}.%
\label{b definition}
\end{equation}

If $\alpha _{G }=1$, this is merely a notational change. However,
$\alpha _{G }$ is a dimensionless parameter that must be determined
by theory or experiment.\ \ We take the independent dimensional
constants to be $(c,\hbar ,b)$\ \ in which case $G=\alpha _{G }\frac{c^{4}}{b}$.\ \ 

\subsubsection{Transformation Equations}

The transformations leaving the symplectic form invariant are $\mathcal{S}p(
4) $ and the transformations leaving the orthogonal metric invariant
are $\mathcal{O}( 2,2) $.\ \ The group of transformations leaving
both invariant is 
\[
\mathcal{U}( 1,1) \simeq \mathcal{U}( 1) \otimes \mathcal{S}\mathcal{U}\left(
1,1\right) \simeq \mathcal{S}p( 4) \cap \mathcal{O}( 2,2) 
\]

Consider first the $\mathcal{S}\mathcal{U}( 1,1) $ transformation
equations 
\begin{equation}
\Xi ( v,f,r) ={\left( 1-w^{2}\right)  }^{-1/2}\left( \begin{array}{llll}
 1  & \frac{v}{ c^{2}} & \frac{f}{ b^{2}} & -\frac{r}{b^{2} c^{2}}
\\
 v & 1 & \frac{r}{b^{2}} & \frac{-f}{ b^{2}} \\
 f & -\frac{r}{c^{2} } & 1 & \frac{v}{c^{2}\ \ } \\
 r & -f &  v  & 1 
\end{array}\right) .%
\label{quaplectic group v f r matrix}
\end{equation}

\noindent The transformation equations are then given\ \ $d\tilde{z}=\Xi
d z$
\begin{gather*}
d\tilde{t}={\left( 1-w^{2}\right)  }^{-1/2}\left( \mathit{dt} +
\frac{v}{c^{2}}\mathit{dq}+ \frac{f}{b^{2}}\mathit{dp}\mathrm{-}\frac{r}{b^{2}
c^{2}}\mathit{de}\right) ,
\end{gather*}
\begin{gather*}
d\tilde{q}={\left( 1-w^{2}\right)  }^{-1/2}\left( \mathit{dq} +
v \mathit{dt}\mathrm{+}\frac{r}{b^{2}}\mathit{dp}-\frac{f}{b^{2}}
\mathit{de} \right) ,
\end{gather*}
\begin{gather*}
d\tilde{p}={\left( 1-w^{2}\right)  }^{-1/2}\left(  \mathit{dp} +f
\mathit{dt}\mathrm{-}\frac{r}{c^{2}}\mathit{dq}+\frac{v}{c^{2}}\mathit{de}
\right) 
\end{gather*}
\[
d\tilde{e}={\left( 1-w^{2}\right)  }^{-1/2}\left(  \mathit{de}-f\mathit{dq}+v
\mathit{dp}+r \mathit{dt}\mathit{\ \ }\right) 
\]

\noindent where\ \ $w^{2}=v^{2}/c^{2}+ f^{2}/b^{2}\mathrm{-}r^{2}/b^{2}
c^{2} $.\ \ It may be directly verified that these equations leave
invariant the symplectic metric (23), ${}^{t}\Xi \cdot \zeta \cdot
\Xi =\zeta $ and Born-Green orthogonal metric\ \ (30),\ \ ${}^{t}\Xi
\cdot \eta \cdot \Xi =\eta $.

The usual special relativity equations take a convenient form when
parameterized by angles and hyperbolic angles. This is true also
in this formulation as described in the Appendix 5.2.

Again, these frames are associated with particles that are simply
curves or trajectories in the space of time, position, momentum
and energy $z= (t,q,p,e)\in \mathbb{P}\simeq \mathbb{R}^{4}$.\ \ \ Each
particle has associated with it an {\itshape observer} with a frame
that is a basis of the cotangent space $\mathit{dz}=(\mathit{dt},\mathit{dq},\mathit{dp},\mathit{de})\in
{T^{*}}_{z}\mathbb{P} $. Consider two observers with trajectories\ \ $c:\mathbb{R}\rightarrow
\mathbb{P}: s\mapsto z=c( s) $\ \ and $\tilde{c}:\mathbb{R}\rightarrow
\mathbb{P}: s\mapsto \tilde{z}=\tilde{c}( s) $ that pass in the
neighborhood of each other at some point in $\mathbb{P}$. In this
case, we assume that their frames are related by a rate of change
of position, $v$, a rate of change of momentum, $f$ and\ \ a rate
of change\ \ of energy $r$ with time $t$.\ \ These generalize the
usual transformations of special relativity where it is assumed
that the relative rate of change of momentum and energy are zero.
The transformations no longer have position-time and momentum-energy
invariant subspaces but mix all the degrees of freedom. Thus the
{\itshape dilation} and {\itshape contraction} concepts of special
relativity are generalized to the full space. 

These transformations no longer have position-time as an invariant
subspace. The position-time subspace of time-position-momentum-energy
space is observer frame dependent. Extreme noninertial frames of
very strongly interacting particles relative to the scale $b$ have
energy and momentum degrees of freedom of particle states are transforming
into position and time degrees of freedom of particle states. 

To provide heuristic plausibility that these degrees of freedom
could {\itshape mix} in the manner described,\ \ consider this in
the context of the very early universe where the energy and momentum
degrees of freedom where huge and position and time degrees of freedom
were very small. We are now in a universe where the energy and momentum
degrees of freedom of particle states are generally relatively small
but the position and time degrees of freedom are very large relative
to these scales.

The tangent vectors are
\begin{gather}
\frac{d\tilde{q}}{d\tilde{t}}=\left( \frac{\mathit{dq}}{\mathit{dt}}
+ v\mathrm{+}\frac{r}{b^{2}}\frac{\mathit{dp}}{\mathit{dt}}-\frac{f}{b^{2}}
\frac{\mathit{de}}{\mathit{dt}} \right) /\left( 1+ \frac{v}{c^{2}}\frac{\mathit{dq}}{\mathit{dt}}+
\frac{f}{b^{2}}\frac{\mathit{dp}}{\mathit{dt}}\mathrm{-}\frac{r}{b^{2}
c^{2}}\frac{\mathit{de}}{\mathit{dt}}\right) ,
\end{gather}
\begin{gather}
\frac{d\tilde{p}}{d\tilde{t}}=\left( \frac{\mathit{dp}}{\mathit{dt}}
+f \mathrm{-}\frac{r}{c^{2}}\frac{\mathit{dq}}{\mathit{dt}}+\frac{v}{c^{2}}\frac{\mathit{de}}{\mathit{dt}}
\right) /\left( 1+ \frac{v}{c^{2}}\frac{\mathit{dq}}{\mathit{dt}}+
\frac{f}{b^{2}}\frac{\mathit{dp}}{\mathit{dt}}\mathrm{-}\frac{r}{b^{2}
c^{2}}\frac{\mathit{de}}{\mathit{dt}}\right) ,
\end{gather}
\begin{equation}
\frac{d\tilde{e}}{d\tilde{t}}=\left(  \frac{\mathit{de}}{\mathit{dt}}-f\frac{\mathit{dq}}{\mathit{dt}}+v
\frac{\mathit{dp}}{\mathit{dt}}+r  \right) /\left( 1+ \frac{v}{c^{2}}\frac{\mathit{dq}}{\mathit{dt}}+
\frac{f}{b^{2}}\frac{\mathit{dp}}{\mathit{dt}}\mathrm{-}\frac{r}{b^{2}
c^{2}}\frac{\mathit{de}}{\mathit{dt}}\right) 
\end{equation}

With the identification $v\simeq \frac{\mathit{dq}}{\mathit{dt}}$,
$\frac{\mathit{df}}{\mathit{dt}}\simeq f$ and $\frac{\mathit{de}}{\mathit{dt}}\simeq
r$ this leads to the rate of change of position, momentum and energy
relativity transformation laws
\begin{gather*}
\tilde{\tilde{v}}=\left( \tilde{v} + v\mathrm{+}\frac{r\tilde{f}}{b^{2}}-\frac{f\tilde{r}}{b^{2}}
\right) /\left( 1+ \frac{v\tilde{v}}{c^{2}}+ \frac{f\tilde{f}}{b^{2}}\mathrm{-}\frac{r
\tilde{r}}{b^{2} c^{2}}\right) ,
\end{gather*}
\begin{gather*}
\tilde{\tilde{f}}=\left( \tilde{f} +f \mathrm{-}\frac{r \tilde{v}
}{c^{2}}+\frac{v \tilde{r}}{c^{2}} \right) /\left( 1+ \frac{v\tilde{v}}{c^{2}}+
\frac{f\tilde{f}}{b^{2}}\mathrm{-}\frac{r \tilde{r}}{b^{2} c^{2}}\right)
,
\end{gather*}
\[
\tilde{\tilde{r}}=\left(  \tilde{r}\mathit{-}f\tilde{v} +v\tilde{f}+r
\right) /\left( 1+ \frac{v\tilde{v}}{c^{2}}+ \frac{f\tilde{f}}{b^{2}}\mathrm{-}\frac{r
\tilde{r}}{b^{2} c^{2}}\right) 
\]

Again as in the usual special relativity case, these are bounded.
If $v=\tilde{v}=c$, $f=\tilde{f}=b$ and $r =\tilde{r}= b c$, then
\begin{gather*}
\tilde{\tilde{v}}=\left( \mathit{c} + c\mathrm{+}\frac{b c b}{b^{2}}-\frac{b
b c}{b^{2}}  \right) /\left( 1+ \frac{c c}{c^{2}}+ \frac{b b}{b^{2}}\mathrm{-}\frac{b
c b c}{b^{2} c^{2}}\right) =c
\end{gather*}
\begin{gather*}
\tilde{\tilde{f}}=\left( \mathit{b}+b\mathrm{-}\frac{b c c}{c^{2}}+\frac{c
b c}{c^{2}} \right) /\left( 1+ \frac{c c}{c^{2}}+ \frac{b b}{b^{2}}\mathrm{-}\frac{b
c b c}{b^{2} c^{2}}\right) =b
\end{gather*}
\[
\tilde{\tilde{r}}=\left(  b c -b\mathit{c}+c b+b c  \right) /\left(
1+ \frac{c c}{c^{2}}+ \frac{b b}{b^{2}}\mathrm{-}\frac{b c b c}{b^{2}
c^{2}}\right) = b c
\]

In this formulation, rates of change of position, momentum and energy
with respect to time are all relative. Forces are only meaningful
between particle states. There is no absolute inertial frame or
absolute rest frame.\ \ Position-time, or as we usually say, space-time
itself is relative.

That this identification is correct may be verified by computing
the matrix multiplication for the four dimensional matrix realization
of the group and verifying that it is precisely the above transformation
rules in (35) that are required for the matrix\ \ identity
\begin{equation}
\Xi ( \tilde{\tilde{v}},\tilde{\tilde{f}},\tilde{\tilde{r}}) = \Xi
( \tilde{v},\tilde{f},\tilde{r}) \cdot \Xi ( v,f,r) 
\end{equation}

\noindent to be satisfied. This is exactly the same reasoning as
in the usual special velocity relativity case given in (3)-(5).

\subsubsection{The $\mathcal{U}( 1) $ transformations}

The full group $\mathcal{U}( 1,1) $ of transformations leaving both
the symplectic and orthogonal metrics invariant includes the $\mathcal{U}(
1) $ subgroup. This group commutes with the $\mathcal{S}\mathcal{U}(
1,1) $ transformations which enables us to consider it separately.
The transformation equations for this group are 
\begin{equation}
\begin{array}{l}
 d\tilde{t}= \cos  \theta  \mathit{dt}-\frac{1}{b c}\ \ \sin  \theta
\mathit{de}, \\
 d\tilde{q}= \cos  \theta  \mathit{dq}-\frac{c }{b }\ \ \sin  \theta
\ \ \mathit{dp}, \\
 d\tilde{p}= \cos  \theta  \mathit{dp}+\frac{b }{c }\ \ \sin  \theta
\mathit{dq}\mathit{,} \\
 d\tilde{e}= \cos  \theta  \mathit{de}+ b c\ \ \sin  \theta  d t.
\end{array}%
\label{u1 transformations}
\end{equation}

\noindent Setting $\tan  \theta = \frac{a}{b c}$, the matrix realization
for this is 
\begin{equation}
\Xi \mbox{}^{\circ}( a) \simeq {\left( 1+{\left( \frac{a}{b c}\right)
}^{2}\right) }^{-1/2}\left( \begin{array}{llll}
 1  & 0 & 0 & \frac{-a}{b^{2} c^{2}} \\
 0 & 1 & \frac{-a}{b^{2}} & 0 \\
 0 & \frac{a}{c^{2} } & 1 & 0 \\
 a & 0 & 0  & 1 
\end{array}\right) 
\end{equation}

\noindent and the full $\mathcal{U}( 1,1) $ matrix realization are
\begin{equation}
\Xi ( v,f,r,a) =\Xi \mbox{}^{\circ}( a) \Xi ( v,f,r) %
\label{u11 matrix}
\end{equation}

\noindent The full $\mathcal{U}( 1,1) $ transformation equations
are 
\begin{equation}
d\tilde{z}=\Xi ( v,f,r,a) \cdot d z%
\label{u11 transformation equations}
\end{equation}

The $\mathcal{U}( 1) $ term does not appear in the classical theory
and the transformation equations are therefore restricted to the
$\mathcal{S}\mathcal{U}( 1,1) $ case.\ \ However, what is quite
remarkable, is that they appear in an essential way in the nonabelian
theory that we shall consider shortly. 

\subsubsection{Limiting forms}

As we have noted above $\Xi ( v,0,0) =\Lambda ( v) $ and so in this
special case, the reciprocal relativistic transformation equations
reduce to the usual special relativity equations.\ \ \ \ 
\begin{equation}
\Xi ( v,0,0) =\Lambda ( v) ={\left( 1-{\left( \frac{v}{c}\right)
}^{2}\right) }^{-1/2}\left( \begin{array}{llll}
 1  & \frac{v}{c^{2}} & 0 & 0 \\
  v  & 1  & 0 & 0 \\
 0 & 0 & 1  & \frac{v}{c^{2}} \\
 0 & 0 & v  & 1 
\end{array}\right) %
\label{reciprical relativity matrix to special relativity}
\end{equation}

This is a realization of $\mathcal{S}\mathcal{O}( 1,1) $ subgroup
on this 4 dimensional space.\ \ \ Note also that\ \ 
\begin{equation}
\Xi ( 0,f,0) =\Lambda ^{\prime }( f) ={\left( 1-{\left( \frac{f}{b}\right)
}^{2}\right) }^{-1/2}\left( \begin{array}{llll}
 1  & 0 & \frac{f}{b^{2}} & 0 \\
 0 & 1  & 0 & \frac{-f}{b^{2}} \\
 f & 0 & 1  & 0 \\
 0 & -f & 0 & 1 
\end{array}\right) %
\label{reciprical relativity matrix to force}
\end{equation}

\noindent This is also a realization of a different $\mathcal{S}\mathcal{O}(
1,1) $ subgroup on this 4 dimensional space.\ \ It acts on the $(t,p)$
and $(e,q)$ subspaces. 

Clearly, arranging for $v=r=0$ with $f$ non-zero will only happen
at isolated points on the trajectory. Think of an oscillator describing
a circle in $(p,q)$ space and a corkscrew when you add in time.\ \ There
are points\ \ on this trajectory where\ \ $f=r=0$ with $v$ non-zero
and also points with for $v=r=0$ with $f$ non-zero.\ \ The special
relativity case (41) applies for the first case and the reciprocally
conjugate case (42) for the second.\ \ \ Note that this latter subgroup
leaves invariant the line elements 
\[
-{\mathit{dt}}^{2}+\frac{1}{b^{2}}{\mathit{dp}}^{2},\ \ \ \ \ \ \ \ \ \ {\mathit{dq}}^{2}-\frac{1}{b^{2}}{\mathit{de}}^{2}
\]

Note that these equations reduce as required to the nonrelativistic
equations described in Section 2.1.\ \ In particular, (32) reduces
in the limit to 
\begin{equation}
\begin{array}{rl}
 \operatorname*{\lim }\limits_{b,c\,\rightarrow \:\infty }\Xi (
v,f,r)  & =\operatorname*{\lim }\limits_{b,c\,\rightarrow \:\infty
}{\left( 1-w^{2}\right)  }^{-1/2}\left( \begin{array}{llll}
 1  & \frac{v}{ c^{2}} & \frac{f}{ b^{2}} & -\frac{r}{b^{2} c^{2}}
\\
 v & 1 & \frac{r}{b^{2}} & \frac{-f}{ b^{2}} \\
 f & -\frac{r}{c^{2} } & 1 & \frac{v}{c^{2}\ \ } \\
 r & -f &  v  & 1 
\end{array}\right)  \\
  & =\left( \begin{array}{llll}
 1  & 0 & 0 & 0 \\
 v & 1 & 0 & 0 \\
 f & 0 & 1 & 0 \\
 r & -f &  v  & 1 
\end{array}\right) =\Phi ( v,f,r) 
\end{array}
\end{equation}

\noindent This is just the expression in (22). The rate of change
of position, momentum and energy in (35) in the limit $b,c\rightarrow
\infty $\ \ are simply 
\[
\begin{array}{l}
 \tilde{\tilde{v}}=\tilde{v} + v  \\
 \tilde{\tilde{f}}=\tilde{f}+ f\ \  \\
 \tilde{\tilde{r}}= \tilde{r} + v \tilde{f} - f \tilde{v}\ \ +r
\end{array}
\]

\noindent These are precisely the equations given in (26).

We can also look at the case where rates of change of momentum is
small with respect to $b$ and the rate of change of energy is small
with respect to $b c$. This is the limit $b\rightarrow \infty $.\ \ In
this case, we have 
\begin{equation}
\begin{array}{rl}
 \Upsilon ( v,f,r)  & =\operatorname*{\lim }\limits_{b\,\rightarrow
\:\infty }\Xi ( v,f,r)  \\
  & ={\left( 1-{\left( \frac{v}{c}\right) }^{2}\right) }^{-1/2}\left(
\begin{array}{llll}
 1  & \frac{v}{ c^{2}} & 0 & 0 \\
 v & 1 & 0 & 0 \\
 f & -\frac{r}{c^{2} } & 1 & \frac{v}{c^{2}\ \ } \\
 r & -f &  v  & 1 
\end{array}\right) 
\end{array}%
\label{quaplectic matrix limit b}
\end{equation}

\noindent The rate of change of position, momentum and energy are
given by
\begin{equation}
\begin{array}{l}
 \tilde{\tilde{v}}=\left( \tilde{v} + v  \right) /\left( 1+ \frac{v\tilde{v}}{c^{2}}\right)
, \\
 \tilde{\tilde{f}}=\left( \tilde{f} +f \mathrm{+}\frac{1 }{c^{2}}\left(
r \tilde{v} -v \tilde{r}\right)  \right) /\left( 1+ \frac{v\tilde{v}}{c^{2}}\right)
, \\
 \tilde{\tilde{r}}=\left(  \tilde{r}\mathrm{+}r-f\tilde{v} +v\tilde{f}
\right) /\left( 1+ \frac{v\tilde{v}}{c^{2}}\right) 
\end{array}
\end{equation}

\noindent It can also be verified that this defines a matrix group.\ \ That
is\ \ $\Upsilon ( \tilde{v} ,\tilde{f} ,\tilde{r} ) \cdot \Upsilon
( v,f,r) =\Upsilon ( \tilde{\tilde{v}} ,\tilde{\tilde{f}} ,\tilde{\tilde{r}}
) $ with the addition law defined in (44). Clearly $\Lambda ( v)
=\Upsilon ( v,0,0) $ is a subgroup. 

The action of\ \ $\Xi $ on ${T^{*}}_{z}\mathbb{P}$, the cotangent
vector space spanned by the basis $\mathit{dz}=\{\mathit{dt},\mathit{dq},\mathit{dp},
\mathit{de}\}$ has no invariant subspaces. The transformations {\itshape
mix} all four of the degrees of freedom. Thus it is no longer meaningful
to talk about a\ \ position-time or {\itshape space-time} manifold.\ \ \ In
the limit $b\rightarrow \infty $, corresponding to the physical
case of arbitrary velocities but relatively small rates of change
of momentum relative to $b$ and rates of change of energy relative
to $b c$,\ \ the subspace spanned by $\{\mathit{dt},\mathit{dq}\mathit{\}}$
is invariant. In this case, all observers have a common view of
position-time space, and the\ \ usual concept of {\itshape space-time
}is meaningful.\ \ In the limit $b,c\rightarrow \infty $, corresponding
to relatively small rates of change of position, momentum and energy,
there are three invariant subspaces, $\{\mathit{dt}\mathit{\}}$,
$\{\mathit{dt},\mathit{dq}\mathit{\}}$\ \ and $\{\mathit{dt},\mathit{dp}\mathit{\}}$.\ \ This
means that in the classical limit there is an absolute notion of
time on which all observers agree upon\ \ and also all observers
agree on the\ \ position-time subspace and the momentum-time subspace.\ \ \ 

\subsubsection{Lie Algebra}

The Lie algebra may be computed directly as in the nonrelativistic
case using\ \ \ \ $d \Xi ( v,f,r,a) |_{0}=d v K+d f N+d r M+ d a
U$
\begin{equation}
\begin{array}{ll}
 K=\left( \begin{array}{llll}
 0 & \frac{1}{c^{2}} & 0 & 0 \\
 1 & 0 & 0 & 0 \\
 0 & 0 & 0 & \frac{1}{c^{2}} \\
 0 & 0 & 1 & 0
\end{array}\right) , & N=\left( \begin{array}{llll}
 0 & 0 & \frac{1}{b^{2}} & 0 \\
 0 & 0 & 0 & \frac{-1}{b^{2}} \\
 1 & 0 & 0 & 0 \\
 0 & -1 & 0 & 0
\end{array}\right)  \\
 M=\left( \begin{array}{llll}
 0 & 0 & 0 & \frac{-1}{c^{2}b^{2}} \\
 0 & 0 & \frac{1}{b^{2}} & 0 \\
 0 & \frac{-1}{c^{2}} & 0 & 0 \\
 1 & 0 & 0 & 0
\end{array}\right) , & U=\left( \begin{array}{llll}
 0 & 0 & 0 & \frac{-1}{b^{2} c^{2}} \\
 0 & 0 & \frac{-1}{b^{2}} & 0 \\
 0 & \frac{1}{c^{2} } & 0 & 0 \\
 1 & 0 & 0  & 0 
\end{array}\right) 
\end{array}%
\label{Quaplectic algebra matrices}
\end{equation}

\noindent with
\[
K=\frac{\partial \Xi }{\partial v},\ \ N=\frac{\partial \Xi }{\partial
f}, M=\frac{\partial \Xi }{\partial r}, U=\frac{\partial \Xi }{\partial
a}
\]

\noindent These may also be viewed as an abstract Lie algebra satisfying
the\ \ commutation relations
\begin{equation}
\begin{array}{l}
 \left[ K,N\right] =2 M,\ \ \left[ M,N\right] =2 K,\ \ \left[ M,K\right]
=-2 N \\
 \left[ \mathit{U},N\right] =0,\ \ \left[ \mathit{U},K\right] =0,\ \ \left[
\mathit{U},M\right] =0, 
\end{array}
\end{equation}

\noindent Again, the limiting forms 
\[
F=\operatorname*{\lim }\limits_{b\,\rightarrow \:\infty }N ,\ \ \hat{\mathit{M}}=\operatorname*{\lim
}\limits_{b\,\rightarrow \:\infty }M
\]

\noindent satisfy the algebra
\begin{equation}
\left[ K,F\right] =2\hat{\mathit{M}},\ \ \left[ \hat{\mathit{M}},F\right]
=0,\ \ \left[ \hat{\mathit{M}},K\right] =-2 N
\end{equation}

\noindent and the nonrelativistic case is 
\[
G=\operatorname*{\lim }\limits_{c\,\rightarrow \:\infty }K ,\ \ R=\operatorname*{\lim
}\limits_{b,c\,\rightarrow \:\infty }M =\operatorname*{\lim }\limits_{b,c\,\rightarrow
\:\infty }U
\]

\noindent with the corresponding algebra given in (29).
\begin{equation}
\left[ G,F\right] =2 R,\ \ \left[ R,F\right] =0,\ \ \left[ R,G\right]
=0
\end{equation}

\subsubsection{Dimensions}

The theory presented in the preceding section describes the behavior
as the rate of change of momentum approaches a constant $b$ and
the rate of change of energy approaches $c b$ in addition to the
usual special relativity behavior as the rate of change of position
approaches $c$.\ \ The constant $b$ is a new universal physical
constant. It may be defined in terms of the existing constants through
a dimensionless parameter $\alpha _{G}=\frac{b G}{c^{4}}$ as defined
in (31). Just as we can define Planck scales of time, position,
momentum and energy in terms of $\{c,\hbar ,G\}$, we can define
them in terms of $\{c,\hbar ,b\}$as\ \ 
\begin{equation}
\lambda _{t}=\sqrt{\frac{\hbar }{b c}}, \lambda _{q}=\sqrt{\frac{\hbar
c}{b}}, \lambda _{p}=\sqrt{\frac{\hbar  b}{ c}}, \lambda _{e}=\sqrt{\hbar
b c}
\end{equation}

\noindent The relationship between the dimensional scales may be
conveniently represented in a quad.
\begin{equation}
\begin{array}{lllll}
 \lambda _{t} & \leftarrow  & c=\lambda _{q}/\lambda _{t} & \rightarrow
& \lambda _{q} \\
 \uparrow  & \nwarrow  &   & \nearrow  & \uparrow  \\
 b=\lambda _{p}/\lambda _{t} &   & \hbar =\lambda _{t}\lambda _{e}=\lambda
_{q}\lambda _{p} &   & b=\lambda _{e}/\lambda _{q} \\
 \downarrow  & \swarrow  &   & \searrow  & \downarrow  \\
 \lambda _{p} & \leftarrow  & c=\lambda _{e}/\lambda _{p} & \rightarrow
& \lambda _{e}
\end{array}%
\label{dimesion quad}
\end{equation}

\noindent If $\alpha _{G}=1$, then this is simply a rewriting of
the usual Planck scales defined in terms of $\{c, \hbar , G\}$.\ \ The
basis of the cotangent space $\{\mathit{dt},\mathit{dq},\mathit{dp},\mathit{de}\}$\ \ and
the parameters $\{v,f,r\}$ may be made dimensionless simply by dividing
by these scales\ \ 
\begin{equation}
\left\{ d\check{t},d\check{q},d\check{p},d\check{e}\right\} =\left\{
\frac{1}{\lambda _{t}}\mathit{dt},\frac{1}{\lambda _{q}}\mathit{dq},\frac{1}{\lambda
_{p}}\mathit{dp},\frac{1}{\lambda _{e}}\mathit{de}\right\}  
\end{equation}

\noindent and
\begin{equation}
\left\{ \check{v},\check{f},\check{r}\right\} =\left\{ \frac{1}{c}v,\frac{1}{b}f,\frac{1}{b
c}r\right\} 
\end{equation}

\noindent Clearly\ \ $\check{v}=1$ when $v=c$ and likewise $\check{f}=1$
when $f=b$ and $\check{r}=1$ when $r=c b$.\ \ \ \ In terms of these
dimensionless basis, all of the $c, b$ constants that appear in
the preceding section may be eliminated by being simply set to 1.\ \ For
example,\ \ (32)\ \ becomes
\begin{equation}
\begin{array}{l}
 d\check{\tilde{t}}={\left( 1-{\check{w}}^{2}\right)  }^{-1/2}\left(
d\check{t} +\check{ v} \mathit{d}\check{q}+ \check{f}\mathit{d}\check{p}\mathrm{-}\check{r}\mathit{d}\check{e}\right)
, \\
 d\check{\tilde{q}}={\left( 1-{\check{w}}^{2}\right)  }^{-1/2}\left(
d\check{q} + \check{ v}\ \ \mathit{d}\check{t} \mathrm{+}\check{r}\mathit{d}\check{p}-
\check{f} \mathit{d}\check{e} \right) , \\
 d\check{\tilde{p}}={\left( 1-{\check{w}}^{2}\right)  }^{-1/2}\left(
d\check{p} + \check{f} \mathit{d}\check{t} \mathrm{+}\check{r}\mathit{d}\check{q}+\check{
v} \mathit{d}\check{e} \right)  \\
 d\check{\tilde{e}}={\left( 1-{\check{w}}^{2}\right)  }^{-1/2}\left(
d\check{e}\mathit{-}\check{f}\mathit{d}\check{q}+\check{ v} \mathit{d}\check{p}+\check{r}
\mathit{d}\check{t}\mathit{\ \ \ }\right) 
\end{array}%
\label{Finite velocity force boosts}
\end{equation}

\noindent with $\check{w}={\check{ v}}^{2} +{\check{f} }^{2}-{\check{r}}^{2}$.\ \ This
is true for all the equations with respect to the natural scales.
In this sense, the constants $c,b$,\ \ and as we will see, $\hbar
$ are no different than the constants we would have to introduce
if we defined the scales in the $'x'$ position direction to be feet
and the $'y'$ position direction to be meters.\ \ Using dimensionless
quantities defines in terms of these constants eliminates their
appearance in all the equations.\ \ \ In particular, the symplectic
and orthogonal metrics are simply
\[
\begin{array}{l}
 {\mathit{ds}}^{2}\mathit{=}{}^{t} d z \cdot \eta \cdot  d z=-d{\check{t}
}^{\mathrm{2}}+\mathit{d}{\check{q}}^{\mathrm{2}}+\mathit{d}{\check{p}}^{\mathrm{2}}-\mathit{d}{\check{e}}^{\mathrm{2}}
\\
 {}^{t} d z \cdot \zeta \cdot  d z=-\mathit{d}\check{e}\wedge \mathit{d}\check{t}
+\mathit{d}\check{p}\wedge \mathit{d}\check{q}
\end{array}
\]

\section{Inhomogeneous group with Heisenberg nonabelian `translations'
}

\subsection{Translation group and inhomogeneous groups}

The discussion up until this point has been concerned with the homogeneous
group acting as a transformation group on the co-tangent space.\ \ We
can cast these transformations in purely group theoretic terms by
introducing the notions of a semidirect product group. 

Consider a Lie group $\mathcal{G}$, a normal closed subgroup $\mathcal{N}\subset
\mathcal{G}$ has the property that, for all $g\in \mathcal{G}$ and
$n\in \mathcal{N}$, that $g^{-1}\cdot n\cdot g\in \mathcal{N}$.
If in addition there is another closed subgroup $\mathcal{K}\subset
\mathcal{G}$ with $\mathcal{K}\cup \mathcal{N}=\mathcal{G}$ and
$\mathcal{N}\cap \mathcal{K}=e$ where $e$ is the trivial group containing
only the identity, then\ \ $\mathcal{G}$ is the semidirect product
group $\mathcal{G}=\mathcal{K}\otimes _{s}\mathcal{N}$.\ \ Note
that as the automorphism group $\mathcal{A}ut( \mathcal{N}) $ of
$ \mathcal{N}$ is the group of elements with the property that for
$n\in \mathcal{N}$,\ \ $g^{-1}\cdot n\cdot g\in \mathcal{N}$.\ \ Clearly
$\mathcal{G}\subseteq \mathcal{A}ut( \mathcal{N}) $.

The group $\mathcal{N}$ has an algebra that may be identified with
the Lie algebra valued one forms $ d n\in \text{\boldmath $a$}(
\mathcal{N}) \simeq T_{e}\mathcal{N}$.\ \ The group $k\in \mathcal{K}$
acts this algebra through the adjoint action $k^{-1}\cdot d n\cdot
k\in \text{\boldmath $a$}( \mathcal{N}) $.\ \ If\ \ $k= e ^{d k}$,
then the infinitesimal transformations are given by\ \ \ $d \tilde{n}=d
n+[d k,d n]$.\ \ \ 

The four dimensional translation group $\mathcal{T}( 4) $ may be
realized as a $5\times 5$ matrix group.\ \ \ 
\begin{equation}
\mathrm{T}( \check{t},\check{q},\check{p},\check{e}) =\left( \begin{array}{lllll}
 1 & 0 & 0 & 0 & \check{t} \\
 0 & 1 & 0 & 0 & \check{q} \\
 0 & 0 & 1 & 0 & \check{p} \\
 0 & 0 & 0 & 1 & \check{e} \\
 0 & 0 & 0 & 0 & 1
\end{array}\right) 
\end{equation}

\noindent The algebra is given by 
\begin{equation}
\begin{array}{rl}
 \mathit{d}\mathrm{T}( \check{t},\check{q},\check{p},\check{e})
& =d\check{t}T+d\check{q}Q+d\check{p}P+d\check{e}E \\
  & =\frac{1}{\lambda _{t}}d t T+\frac{1}{\lambda _{q}}d q Q+\frac{1}{\lambda
_{p}}d p P+\frac{1}{\lambda _{e}}d e E
\end{array}
\end{equation}

\noindent where the basis $\{T,Q,P,E\}$ of the abelian algebra\ \ $\text{\boldmath
$a$}( \mathcal{T}( 4) ) $ have the matrix realizations 
\begin{equation}
\begin{array}{ll}
 T=\left( \begin{array}{lllll}
 0 & 0 & 0 & 0 & 1 \\
 0 & 0 & 0 & 0 & 0 \\
 0 & 0 & 0 & 0 & 0 \\
 0 & 0 & 0 & 0 & 0 \\
 0 & 0 & 0 & 0 & 0
\end{array}\right)  & Q=\left( \begin{array}{lllll}
 0 & 0 & 0 & 0 & 0 \\
 0 & 0 & 0 & 0 & 1 \\
 0 & 0 & 0 & 0 & 0 \\
 0 & 0 & 0 & 0 & 0 \\
 0 & 0 & 0 & 0 & 0
\end{array}\right)  \\
 P=\left( \begin{array}{lllll}
 0 & 0 & 0 & 0 & 0 \\
 0 & 0 & 0 & 0 & 0 \\
 0 & 0 & 0 & 0 & 1 \\
 0 & 0 & 0 & 0 & 0 \\
 0 & 0 & 0 & 0 & 0
\end{array}\right)  & E=\left( \begin{array}{lllll}
 0 & 0 & 0 & 0 & 0 \\
 0 & 0 & 0 & 0 & 0 \\
 0 & 0 & 0 & 0 & 0 \\
 0 & 0 & 0 & 0 & 1 \\
 0 & 0 & 0 & 0 & 0
\end{array}\right) 
\end{array}%
\label{Translation generators}
\end{equation}

\noindent More compactly, with $\{Z_{\alpha }\} =\{T,Q,P,E\}$ and
$\{{\mathit{z}}^{\alpha }\}=\{\check{t},\check{q},\check{p},\check{e}\mathit{\}}$,
\[
d \mathrm{T}( z) = d z^{\alpha }Z_{\alpha }=\left( \begin{array}{ll}
 0 & d z \\
 0 & 0
\end{array}\right)  
\]

Consider the semidirect product group\ \ $\mathcal{G}=\mathcal{K}\otimes
_{s}\mathcal{T}( 4) $.\ \ The automorphism group of $\mathcal{T}(
n) $ is $\mathcal{G}\mathcal{L}( n) \otimes _{s}\mathcal{T}( n)
$ and therefore $\mathcal{K}\subseteq \mathcal{G}\mathcal{L}( 4)
$.\ \ Therefore elements $k\in \mathcal{K}$ may be realized by nonsingular
$4\times 4$ matrices $K$ and elements $g\in \mathcal{G}$ may be
realized by the\ \ $5\times 5$ matrices $\Gamma $ 
\begin{equation}
 \Gamma ( K,z) = \left( \begin{array}{ll}
 K & K\cdot z \\
 0 & 1
\end{array}\right) 
\end{equation}

\noindent with group product $\Gamma ( \tilde{K},\tilde{\mathit{z}})
\cdot \Gamma ( K,z) =\Gamma ( \tilde{K}\cdot K,\tilde{K}\cdot z+\tilde{\mathit{z}})
$ $\ \ $and inverse $\Gamma ^{-1}( K,z) =\Gamma ( K^{-1},-z) $.

\subsubsection{Lie algebra}

\noindent Now, we can consider, in particular, the group $\mathcal{K}\simeq
\mathcal{U}( 1,1) $ and set $\mathrm{K}=\Xi ( v,f,r,a) $ from (39).
The action of $\Xi $ on the element of the algebra $d \mathrm{T}(
z) $ is
\begin{equation}
\begin{array}{rl}
 \Xi  d \mathrm{T}( z) \Xi ^{-1} & =\left( \begin{array}{ll}
 \Xi  & 0 \\
 0 & 1
\end{array}\right) \left( \begin{array}{ll}
 0 & d z \\
 0 & 0
\end{array}\right) \left( \begin{array}{ll}
 \Xi ^{-1} & 0 \\
 0 & 1
\end{array}\right)  \\
  & =\left( \begin{array}{ll}
 0 & \Xi \cdot d z \\
  0 & 0
\end{array}\right)  = \mathit{d} \mathrm{T}( \tilde{z}) 
\end{array}%
\label{translation matrix}
\end{equation}

\noindent This is precisely the transformation equations given in
(32),\ \ \ \ $d \tilde{z}= \Xi \cdot \mathit{dz}$. 

The generators\ \ $\{Z_{\alpha }\}=\{T,Q,P,E\}$ are given in (57)\ \ and
the generators $\{K,N,M,U\}$ of the Lie algebra are given in (46)
with the embedding of the $ $$4\times 4$ matrices in the $5\times
5$\ \ matrices.\ \ The nonzero generators of the\ \ algebra of $\mathcal{U}(
1,1) \otimes _{s}\mathcal{T}( 4) $\ \ are\ \ $d \Xi ( v,f,r,a) |_{0}=d
v K + d f N+d r M+d a U$
\begin{equation}
\begin{array}{llll}
 \left[ K,N\right] =2 M, & \left[ M,N\right] =2 K, & \left[ M,K\right]
=-2 N &   \\
 \left[ \mathit{K},T\right] =Q, & \left[ \mathit{K},Q\right] =T,
& \left[ \mathit{K},P\right] =E, & \left[ \mathit{K},E\right] =P,
\\
 \left[ \mathit{N},T\right] =P, & \left[ \mathit{N},Q\right] =-E,
& \left[ \mathit{N},P\right] =T, & \left[ \mathit{U},E\right] =-Q,
\\
 \left[ \mathit{M},T\right] =E, & \left[ \mathit{M},Q\right] =-P,
& \left[ \mathit{M},P\right] =Q, & \left[ \mathit{M},E\right] =-T,
\\
 \left[ \mathit{U},T\right] =-E, & \left[ \mathit{U},Q\right] =-P,
& \left[ \mathit{U},P\right] =Q, & \left[ \mathit{U},E\right] =T,
\end{array}
\end{equation}

\noindent The infinitesimal transformation equations of (32) are
then given by the action of the algebra of the $\mathcal{S}\mathcal{U}(
1,1) $ subgroup\ \ 
\begin{equation}
\begin{array}{rl}
 d \mathrm{T}( \tilde{z})  & =d \mathrm{T}( z) +\left[ d \Xi ( v,f,r)
|_{0}, d \mathrm{T}( z)  \right]  \\
  & =d z^{\alpha }Z_{\alpha }+d \left. z^{\alpha }  [\mathit{dv}
K+d\mathit{f} \mathit{N}\mathrm{+}d r M ,Z_{\alpha }\right]  \\
  & =d z^{\alpha }( Z_{\alpha }+d \left. v  [K ,Z_{\alpha }\right]
+\left. d\mathit{f} [\mathit{N},Z_{\alpha }\right] +d \left. r [\mathit{M}
,Z_{\alpha }\right] ) 
\end{array}
\end{equation}

\noindent This gives the expected result for the infinitesimal transformations
\begin{equation}
\begin{array}{l}
 d\tilde{t}=\mathit{dt} + \frac{1}{c^{2}}d v \wedge \mathit{dq}+
\frac{1}{b^{2}}d f\wedge  \mathit{dp}\mathrm{-}\frac{1}{b^{2} c^{2}}d
r \wedge \mathit{de}, \\
 d\tilde{q}=\mathit{dq} + d v \wedge \mathit{dt}\mathrm{+}\frac{1}{b^{2}}d
r \wedge \mathit{dp}-\frac{1}{b^{2}}d f \wedge \mathit{de} , \\
 d\tilde{p}= \mathit{dp} +d f \wedge \mathit{dt}\mathrm{-}\frac{1}{c^{2}}d
r \wedge \mathit{dq}+\frac{1}{c^{2}}d v\wedge \mathit{de}\mathrm{,}\text{}
\\
 d\tilde{e}= \mathit{de}-d f\wedge \mathit{dq}+d v\wedge  \mathit{dp}+d
r\wedge  d t\mathit{.} 
\end{array}%
\label{isu11 infinitesimal transformation equations}
\end{equation}

The lowest order Casimir invariant for this algebra is $-T^{2}+Q^{2}+P^{2}-E^{2}$
which is precisely the form of the orthogonal metric. This enables
all the essential properties to be formulated in purely group theoretic
terms. 

\subsection{Heisenberg group}

The theory considered so far does not take into account the nonabelian
nature of phase space.\ \ Physical observations tell us that the
position and momentum degrees of freedom and the time and energy
degrees of freedom cannot be measured simultaneously.\ \ Mathematically,
this means that the abelian group $\mathcal{T}( 4) $ in the previous
section must be replaced by a Heisenberg group $\mathcal{H}( 2)
=\mathcal{T}( 2) \otimes _{s }\mathcal{T}( 3) $.\ \ This is a 5
dimensional group with an algebra that has the generators $\{T,Q,P,E,I\}$
satisfying the Lie algebra
\begin{equation}
\left[ P,Q\right] =-I,\ \ \ \left[ T,E\right] =I%
\label{Heisenberg algebra}
\end{equation}

The group $\mathcal{H}( 2) $ is just a matrix group like all the
groups we have encountered so far. It can be realized by the $6\times
6$\ \ matrices\ \ \ 
\begin{equation}
\mathrm{H}( \check{t},\check{q},\check{p},\check{e},\iota ) =\left(
\begin{array}{llllll}
 1 & 0 & 0 & 0 & 0 & \check{t} \\
 0 & 1 & 0 & 0 & 0 & \check{q} \\
 0 & 0 & 1 & 0 & 0 & \check{p} \\
 0 & 0 & 0 & 1 & 0 & \check{e} \\
 -\check{e} & \check{p} & -\check{q} & \check{t} & 1 & 2 \iota 
\\
 0 & 0 & 0 & 0 & 0 & 1
\end{array}\right) 
\end{equation}

\noindent This may be written more compactly as 
\begin{equation}
\mathrm{H}( z,\iota ) =\left( \begin{array}{lll}
 I & 0 & z \\
 {}^{t}z\cdot \zeta  & 1 & 2 \iota  \\
 0 & 0 & 1
\end{array}\right) %
\label{Heisenberg matrix condensed}
\end{equation}

\noindent It follows that the group product and inverse are\ \ 
\begin{equation}
\mathrm{H}( \tilde{z},\tilde{\iota }) \cdot \mathrm{H}( z,\iota
) =\mathrm{H}( \tilde{z}+z,\tilde{\iota }+\iota +\frac{1}{2}{}^{t}\tilde{z}\cdot
\zeta \cdot z) ,\text{\ \ ${\mathrm{H}( z,\iota ) }^{-1 }=\mathrm{H}(
-z,-\iota ) $}
\end{equation}

\noindent and the algebra is 
\begin{equation}
d \mathrm{H}( z,\iota ) =\left( \begin{array}{lll}
 0 & 0 & d z \\
 {}^{t}\left( \zeta \cdot d z\right)  & 0 & 2 d \iota  \\
 0 & 0 & 0
\end{array}\right) =d z^{\alpha }Z_{\alpha }+d \iota  I
\end{equation}

Expanding out again, this means explicitly that the generators are
given by the $6\times 6$ matrices
\begin{equation}
\begin{array}{ll}
 T=\left( \begin{array}{llllll}
 0 & 0 & 0 & 0 & 0 & 1 \\
 0 & 0 & 0 & 0 & 0 & 0 \\
 0 & 0 & 0 & 0 & 0 & 0 \\
 0 & 0 & 0 & 0 & 0 & 0 \\
 0 & 0 & 0 & -1 & 0 & 0 \\
 0 & 0 & 0 & 0 & 0 & 0
\end{array}\right)  & Q=\left( \begin{array}{llllll}
 0 & 0 & 0 & 0 & 0 & 0 \\
 0 & 0 & 0 & 0 & 0 & 1 \\
 0 & 0 & 0 & 0 & 0 & 0 \\
 0 & 0 & 0 & 0 & 0 & 0 \\
 0 & 0 & 1 & 0 & 0 & 0 \\
 0 & 0 & 0 & 0 & 0 & 0
\end{array}\right)  \\
 P=\left( \begin{array}{llllll}
 0 & 0 & 0 & 0 & 0 & 0 \\
 0 & 0 & 0 & 0 & 0 & 0 \\
 0 & 0 & 0 & 0 & 0 & 1 \\
 0 & 0 & 0 & 0 & 0 & 0 \\
 0 & -1 & 0 & 0 & 0 & 0 \\
 0 & 0 & 0 & 0 & 0 & 0
\end{array}\right)  & E=\left( \begin{array}{llllll}
 0 & 0 & 0 & 0 & 0 & 0 \\
 0 & 0 & 0 & 0 & 0 & 0 \\
 0 & 0 & 0 & 0 & 0 & 0 \\
 0 & 0 & 0 & 0 & 0 & 1 \\
 1 & 0 & 0 & 1 & 0 & 0 \\
 0 & 0 & 0 & 0 & 0 & 0
\end{array}\right)  \\
 I=\left( \begin{array}{llllll}
 0 & 0 & 0 & 0 & 0 & 0 \\
 0 & 0 & 0 & 0 & 0 & 0 \\
 0 & 0 & 0 & 0 & 0 & 0 \\
 0 & 0 & 0 & 0 & 0 & 0 \\
 0 & 0 & 0 & 0 & 0 & 2 \\
 0 & 0 & 0 & 0 & 0 & 0
\end{array}\right)  &  
\end{array}%
\label{Heisenberg algebra matrices}
\end{equation}

These matrices satisfy the algebra of the algebra of the Heisenberg
matrix Lie group. In quantum mechanics where the Heisenberg algebra
normally appears, we are using the unitary representations of the
group on a Hilbert space.\ \ In general an element of a Lie group\ \ $g\in
\mathcal{G}$ is given in the neighborhood of the identity by $g=
e ^{X}$ where $X\in \text{\boldmath $a$}( \mathcal{G}) $ is an element
of the algebra. A unitary representation $\varrho $ of the group
as unitary operators on a Hilbert space ${\text{\boldmath $H$}}^{\varrho
}$. The unitary irreducible representations determine the Hilbert
space and so it is labelled by the representation. The unitary operators\ \ ${\varrho
( g) }^{\dagger }={\varrho ( g) }^{-1}$induce anti-Hermitian representations
$\varrho '{(X)}^{\dagger }=-\varrho '(X)$ of the algebra.\ \ Inserting
an $i$ maps these onto Hermitian operators $\varrho ( g) = e ^{i
\varrho ( g) }$\ \ normally used in quantum mechanics. Then if the
Lie algebra is $[X,Y]=Z$, then the Lie algebra of the Hermitian
representation is $[\varrho '(X),\varrho '(Y)]=-i \varrho '(Z)$.\ \ This
is where the $i$ comes from in the Heisenberg algebra as we generally
deal with the unitary representations in a quantum mechanics context
where the Hilbert space is ${\text{\boldmath $H$}}^{\varrho }=L^{2}(
\mathbb{R},\mathbb{C}) $ and the representations $\hat{Q}=\varrho
'(Q), $$\hat{P}=\varrho '(P)$ and $\hat{I}=\varrho '(I)$ of the
algebra satisfy $[\hat{P},\hat{Q}]= i \hbar \hat{I} $\ \ and are
realized in the position diagonal basis by the Hermitian operators
$\hat{Q}=q$ and $\hat{P}=i \hbar \frac{\partial }{\partial q}$ or
the momentum diagonal basis by\ \ $\hat{Q}=-i \hbar  \frac{\partial
}{\partial p}$ and $\hat{P}=p$ .\ \ 

The Heisenberg group itself, before the unitary representations
are considered,\ \ is just a straightforward real matrix group as
are the rotation, symplectic and translation groups with which we
are familiar. 

\subsection{Automorphisms of the Heisenberg group}\label{automorphism
section}

We can now proceed as in the case of the translation group and construct
a semidirect product group that contains the Heisenberg group as
the normal subgroup. This group must be a subgroup of the automorphism
group of the Heisenberg group. That is for\ \ $a\in \mathcal{A}ut(
\mathcal{H}( 2) ) $,\ \ $a^{-1}\cdot h\cdot a \in \mathcal{H}( 2)
$ for all $h\in \mathcal{H}( 2) $.\ \ Constructing a general $6\times
6$ matrix\ \ and computing the product shows that the automorphism
group must be of the form 
\begin{equation}
\mathcal{A}ut( \mathcal{H}( 2) ) \simeq \mathcal{D}\otimes _{s}\left(
\mathcal{A}( 1) \otimes _{s}\left( \mathcal{S}p( 4) \otimes _{s}\mathcal{H}(
2) \right) \right) 
\end{equation}

\noindent $\mathcal{D}$ is the discrete group of automorphisms that
is considered further in Section 4.5 below.\ \ $\mathcal{A}$ is
the abelian group of dilations that are represented by the\ \ $6\times
6$ matrices\ \ $A( \epsilon ) \in \mathcal{A}$ 
\begin{equation}
A( \epsilon ) = \left( \begin{array}{lll}
 I & 0 & 0 \\
 0 & e^{\epsilon } & 0 \\
 0 & 0 & e^{-\epsilon }
\end{array}\right) 
\end{equation}

\noindent The product is $A( \tilde{\epsilon }) \cdot A( \epsilon
) =A( \tilde{\epsilon }+\epsilon ) $ and inverse ${A( \epsilon )
}^{-1}=A( -\epsilon ) $.\ \ The action as an automorphism of the
Heisenberg group is\ \ 
\begin{equation}
A( \epsilon ) \cdot \mathrm{H}( z,\iota ) \cdot {A( \epsilon ) }^{-1}=\left(
\begin{array}{lll}
 I & 0 &  e ^{\epsilon }z \\
 {}^{t}\left( \zeta \cdot  e ^{\epsilon }z\right)  & 1 & 2  e ^{2\epsilon
}\iota  \\
 0 & 0 & 1
\end{array}\right) =\mathrm{H}(  e ^{\epsilon }z, e ^{2\epsilon
}\iota ) 
\end{equation}

\noindent The remaining automorphisms are the $6\times 6$ matrices\ \ \ $\Upsilon
\in \mathcal{S}p( 4) \otimes \mathcal{H}( 2) $
\begin{equation}
\Upsilon ( K,z,\iota ) =\left( \begin{array}{lll}
 K & 0 & K\cdot  z \\
 {}^{t}z\cdot \zeta  & 1 & 2\ \ \iota  \\
 0 & 0 & 1
\end{array}\right) 
\end{equation}

\noindent where $K\in \mathcal{S}p( 4) $ are the $4\times 4$ matrices
with the property that\ \ ${}^{t}K\cdot \zeta \cdot K=\zeta $.\ \ \ This
property may be used to determine the group multiplication and inverse.\ \ \ As
this is a matrix group, the group product and inverse is computed
directly from the matrix product and inverse is calculated us 
\begin{gather*}
\Upsilon ( \tilde{K},\tilde{z},\tilde{\iota }) \cdot \Upsilon (
K,z,\iota ) =\Upsilon ( \tilde{K}\cdot K,\tilde{z}+\tilde{K}\cdot
z,\iota +\tilde{\iota }+\frac{1}{2}{}^{t}\tilde{z}\cdot \zeta \cdot
z) 
\end{gather*}
\[
{\Upsilon ( K,z,\iota ) }^{-1}= \Upsilon ( K^{-1},-z,-\iota ) 
\]

Finally the automorphisms are 
\begin{equation}
\begin{array}{rl}
 \Upsilon \left( \tilde{K},\tilde{z},\tilde{\iota }\right) \cdot
\mathrm{H}( z,\iota ) \cdot {\Upsilon ( \tilde{K},\tilde{z},\tilde{\iota
}) }^{-1} & =\left( \begin{array}{lll}
 I & 0 & \tilde{K}\cdot z \\
 {}^{t}z\cdot \zeta \cdot {\tilde{K}}^{-1} & 1 & 2\left( \iota +{}^{t}\tilde{z}\cdot
\zeta \cdot z\right)  \\
 0 & 0 & 1
\end{array}\right) 
\end{array}
\end{equation}

\noindent This is an element of $\mathcal{H}( 2) $ only if\ \ ${}^{t}z\cdot
\zeta \cdot {\tilde{K}}^{-1}={}^{t}(\tilde{K}\cdot z)\cdot \zeta
={}^{t}z\cdot {}^{t}\tilde{K}\cdot \zeta $. That is, $\zeta \cdot
{\tilde{K}}^{-1}={}^{t}\tilde{K}\cdot \zeta $ or equivalently $\zeta
={}^{t}\tilde{K}\cdot \zeta \cdot \tilde{K}$.\ \ This is the condition
for $\tilde{K}\in \mathcal{H}( 2) $ and the automorphisms are then
\begin{equation}
\begin{array}{rl}
 \Upsilon \left( \tilde{K},\tilde{z},\tilde{\iota }\right) \cdot
\mathrm{H}( z,\iota ) \cdot {\Upsilon ( \tilde{K},\tilde{z},\tilde{\iota
}) }^{-1} & =\mathrm{H}( \tilde{K}\cdot z,\iota +{}^{t}\tilde{z}\cdot
\zeta \cdot z) 
\end{array}
\end{equation}

\subsection{Quaplectic group}

Then, the group $\mathcal{Q}( 1,1) =\mathcal{U}( 1,1) \otimes _{s}\mathcal{H}(
2) $ is a subgroup of the group of continuous automorphisms with
elements realized by the matrices 
\begin{equation}
\Theta =\left( \begin{array}{lll}
 \Xi  & 0 & \Xi \cdot z \\
 {}^{t}z\cdot \zeta  & 1 & 2 \iota  \\
 0 & 0 & 1
\end{array}\right) 
\end{equation}

\noindent This is an element of the quaplectic group in a matrix
realization. 

The full group including the discrete transformations and the scaling
transformations is the extended quaplectic group $\hat{\mathcal{Q}}(
1,1) =(\mathcal{D}\otimes \mathcal{A}b\otimes \mathcal{U}( 1,1)
)\otimes _{s}\mathcal{H}( 2)  $ with elements realized by the matrices
\cite{folland} 
\begin{equation}
\hat{\Theta }=\varsigma  \left( \begin{array}{lll}
 \Xi  & 0 & \Xi \cdot z \\
 e^{\epsilon }{}^{t}z\cdot \zeta  & e^{\epsilon } & 2 e^{\epsilon
}\iota  \\
 0 & 0 & e^{-\epsilon }
\end{array}\right) =\varsigma \cdot A\cdot \Xi \cdot \mathrm{H}%
\label{extended quaplectic group}
\end{equation}

\noindent where $\varsigma \in \mathcal{D}$ is an element of the
finite abelian discrete\ \ group that is defined in the following
section. 

The quaplectic group gives the transformation equations
\begin{equation}
\begin{array}{rl}
 \Xi \ \ \mathit{d} \mathrm{H}( z) \Xi ^{-1} & = \left( \begin{array}{lll}
 \Xi  & 0 & 0 \\
 0 & 1 & 0 \\
 0 & 0 & 1
\end{array}\right) \left( \begin{array}{lll}
 0 & 0 & d z \\
 {}^{t}d z\cdot \zeta  & 0 & 2 d \iota  \\
 0 & 0 & 0
\end{array}\right) \left( \begin{array}{lll}
 \Xi ^{-1} & 0 & 0 \\
 0 & 1 & 0 \\
 0 & 0 & 1
\end{array}\right)  \\
  & =\left( \begin{array}{lll}
 0 & 0 & \Xi \cdot \mathit{d} \mathit{z} \\
 {}^{t}\left( \Xi \cdot \mathit{d} \mathit{z}\right) \cdot \zeta
& 0 & 2 d \iota  \\
 0 & 0 & 0
\end{array}\right) = \mathit{d} \mathrm{H}( \tilde{z}) 
\end{array}%
\label{Heisenberg translation matrix}
\end{equation}

\noindent where we have used\ \ $ {}^{t}d z\cdot \zeta \cdot \Xi
^{-1}={}^{t}d z\cdot  \mathrm{{}}{}^{t}\Xi \cdot  {}^{t}\Xi ^{-1}\cdot
\zeta \cdot \Xi ^{-1} = {}^{t}(\Xi \cdot \mathit{d} \mathit{z})\cdot
\zeta $\ \ as\ \ \ ${}^{t}\Xi \cdot \zeta \cdot \Xi =\zeta $ as
$\Xi \in \mathcal{S}p( 4) $ and therefore\ \ \ \ ${}^{t}\Xi ^{-1}\cdot
\zeta \cdot \Xi ^{-1}=\zeta $\ \ with $\zeta ^{-1}=-\zeta $.\ \ These
are precisely the transformation equations\ \ \ $d\tilde{z}= \Xi
\cdot \mathit{dz}$ in\ \ (40).\ \ 

\subsubsection{Lie Algebra}

The Lie algebra of the quaplectic group is 
\begin{equation}
\begin{array}{l}
 \mathit{d}\Theta = \mathit{d}\Xi +d z^{\alpha }Z_{\alpha }+\mathit{d}\iota
I \\
 d \Xi =d v K+d f N+d r M+ d a U
\end{array}
\end{equation}

\noindent The generators\ \ $\{Z_{\alpha },I\}=\{T,Q,P,E,I\}$ are
given in (68)\ \ and the generators $\{K,N,M,M \mbox{}^{\circ}\}$
are given in (46) with the obvious embedding of the $4\times 4$
matrices in the $6\times 6$ matrices.\ \ The nonzero generators
of the full quaplectic algebra are 
\begin{equation}
\begin{array}{llll}
 \left[ K,N\right] =2 M, & \left[ M,N\right] =2 K, & \left[ M,K\right]
=-2 N &   \\
 \left[ \mathit{K},T\right] =Q, & \left[ \mathit{K},Q\right] =T,
& \left[ \mathit{K},P\right] =E, & \left[ \mathit{K},E\right] =P,
\\
 \left[ \mathit{N},T\right] =P, & \left[ \mathit{N},Q\right] =-E,
& \left[ \mathit{N},P\right] =T, & \left[ \mathit{U},E\right] =-Q,
\\
 \left[ \mathit{M},T\right] =E, & \left[ \mathit{M},Q\right] =-P,
& \left[ \mathit{M},P\right] =Q, & \left[ \mathit{M},E\right] =-T,
\\
 \left[ \mathit{U},T\right] =-E, & \left[ \mathit{U},Q\right] =-P,
& \left[ \mathit{U},P\right] =Q, & \left[ \mathit{U},E\right] =T,
\\
 \left[ P,Q\right] =-I, & \left[ E,T\right] =I &   &  
\end{array}
\end{equation}

\noindent The infinitesimal transformation equations\ \ of (32)
are then given by 
\begin{equation}
\begin{array}{rl}
 d \mathrm{H}( \tilde{z})  & =d \mathrm{H}( z) \ \ +\left[ d \Xi
( v,f,r,a) |_{0}, d \mathrm{H}( z)  \right]  \\
  & =d z^{\alpha }Z_{\alpha }+\left. {\mathit{dz}}^{\alpha }  [\mathit{dv}
K+d\mathit{f} \mathit{N}\mathrm{+}\mathrm{dr} M\mathit{+}\mathit{d}
\mathit{a}  \mathit{U},Z_{\alpha }\right]  \\
  & =d z^{\alpha }( Z_{\alpha }+d \left. v  [K ,Z_{\alpha }\right]
+\left. d\mathit{f} [\mathit{N},Z_{\alpha }\right] +d \left. r [\mathit{M}
,Z_{\alpha }\right] +d \left. a [\mathit{U} ,Z_{\alpha }\right]
) 
\end{array}
\end{equation}

\noindent This gives the expected result on the non abelian manifold
\begin{equation}
\begin{array}{l}
 d\tilde{t}=\mathit{dt} + \frac{1}{c^{2}}d v \wedge \mathit{dq}+
\frac{1}{b^{2}}d f\wedge  \mathit{dp}\mathrm{-}\frac{1}{b^{2} c^{2}}\left(
d r +d a\right) \wedge \mathit{de}, \\
 d\tilde{q}=\mathit{dq} + d v \wedge \mathit{dt}\mathrm{+}\frac{1}{b^{2}}\left(
d r -d a\right) \wedge \mathit{dp}-\frac{1}{b^{2}}d f \wedge \mathit{de}
, \\
 d\tilde{p}= \mathit{dp} +d f \wedge \mathit{dt}\mathrm{-}\frac{1}{c^{2}}\left(
d r -d a\right) \wedge \mathit{dq}+\frac{1}{c^{2}}d v\wedge  \mathit{de}\mathrm{,}\text{}
\\
 d\tilde{e}= \mathit{de}-d f\wedge \mathit{dq}+d v\wedge  \mathit{dp}+\left(
d r+ d a\right) \wedge  d t\mathit{.} 
\end{array}%
\label{quaplectic infinitesimal transformation equations}
\end{equation}

In this case however, the lowest order Casimir invariant is simply
$C_{1}=I$ as it commutes with all the generators and the second
order Casimir invariant is 
\begin{equation}
C_{2}=\frac{1}{2}\left( -T^{2}-Q^{2}+P^{2}-E^{2} \right) - I U%
\label{Casimir invariant 2}
\end{equation}

\noindent This can be viewed as a metric on a 6 dimensional space
(or with the $n=3$ case, a $10$ dimensional space.) The additional
term is required as the $Q$ and $P$ and the $T$ and $E$ do not commute.\ \ However,
this additional generator associated with the $\mathcal{U}( 1) $
subgroup provides precisely the term to cancel out the resulting
term when considering Lie brackets of the form $[Z_{\alpha },C_{2}]$.\ \ \ That
is\ \ 
\begin{equation}
\left[ T,C_{2}\right]  = -2\frac{1}{2}E[ T,E] -I[ T,U] =-E I+I E=0
\end{equation}

\noindent and so forth. 

This is a very essential change in the structure of the theory.
One of the most profound of these is that the abelian theory with
a hermitian metric does not admit a non-trivial manifold structure
to enable the mathematical development of a general theory on this
space corresponding to the general relativity generalization of
special relativity. This is the no go theorem of Schuller \cite{Schuller}.
However, this nonabelian theory does not appear to be constrained
by this no go theorem and it is possible to investigate generalizations
to general curved noncommutative manifolds.\ \ The construction
of a nonabelian geometry where locally the nonabelian Heisenberg
group are the local\ \ {\itshape translations} that generalize to
a nonabelian connection is a very interesting follow on problem.

\subsection{Discrete transformations}

The parity, time reversal and charge conjugation (PCT) discrete
transformations play an important role in the standard special relativistic
theory.\ \ These transformations\ \ \ 
\begin{equation}
\begin{array}{ll}
 \varsigma _{P}\cdot \left\{ \mathit{dt},\mathit{dq}\right\} =\left\{
\mathit{dt},-\mathit{dq}\right\} , & \varsigma _{T}\cdot \left\{
\mathit{dt},\mathit{dq}\right\} =\left\{ -\mathit{dt},\mathit{dq}\right\}
, \\
 \varsigma _{C}\cdot \left\{ \mathit{dt},\mathit{dq}\right\} =\left\{
-d t,-d q\right\}  &  
\end{array}\ \ %
\label{pct discrete transformations}
\end{equation}

\noindent are a discrete abelian group satisfying ${\varsigma _{P}}^{2}={\varsigma
_{T}}^{2}={\varsigma _{C}}^{2}=\varsigma _{0}$,\ \ $\varsigma _{P}\cdot
\varsigma _{T}=\varsigma _{C}$, $\varsigma _{C}\cdot \varsigma _{P}=\varsigma
_{T}$, $\varsigma _{T}\cdot \varsigma _{C}=\varsigma _{P}$. These
transformations leave the Lorentz metric invariant and are automorphisms
of the group
\begin{equation}
\varsigma _{P}\cdot \Lambda ( v) \cdot {\varsigma _{P}}^{-1}=\Lambda
( -v) ,\ \ \varsigma _{T}\cdot \Lambda ( v) \cdot {\varsigma _{T}}^{-1}=\Lambda
( -v) ,\ \ \varsigma _{C}\cdot \Lambda ( v) \cdot {\varsigma _{C}}^{-1}=\Lambda
( v) 
\end{equation}

\noindent These transformations carry over directly to the $\mathcal{U}(
1,3) $ group of discrete automorphisms $\mathcal{D}\mbox{}^{\circ}$.\ \ Again,
define\ \ $d z=\{\mathit{dt},\mathit{dq},\mathit{dp}, \mathit{de}\}$,
\begin{equation}
\begin{array}{ll}
 \varsigma _{P}\left( \mathit{dz}\right) =\left\{ \mathit{dt},-\mathrm{dq},-\mathit{dp},
\mathit{de}\right\} , & \varsigma _{T}\left( \mathit{dz}\right)
=\left\{ -\mathit{dt},\mathit{dq},\mathit{dp}, -\mathit{de}\right\}
, \\
 \varsigma _{C}\left( \mathit{dz}\right) =\left\{ -\mathit{dt},-\mathit{dq},-\mathit{dp},-\mathit{de}\right\}
&  
\end{array}%
\label{Discrete PCT transformations}
\end{equation}

\noindent These satisfy the above group product relations given
above in (84).\ \ \ There are now in addition 3 new discrete transformations,
the Born reciprocity transformations\ \ \cite{born-einstein}\ \ (using
units $b=c=\hbar =1)$
\begin{equation}
\begin{array}{ll}
 \varsigma _{E}\left( \mathit{dz}\right) =\left\{ -\mathit{de},\mathit{dq},\mathit{dp},
\mathit{dt}\right\}  & \varsigma _{Q}\left( \mathit{dz}\right) =\left\{
\mathit{dt},\mathit{dp},-\mathit{dq}, \mathit{de}\right\} , \\
 \varsigma _{R}\left( \mathit{dz}\right) =\left\{ -\mathit{de},\mathit{dp},-\mathit{dq},\mathit{dt}\right\}
&  
\end{array}%
\label{Discrete QER}
\end{equation}

\noindent The group products for these transformations are ${\varsigma
_{R}}^{2}=\varsigma _{C}$, ${\varsigma _{Q}}^{2}=\varsigma _{P}$,
${\varsigma _{E}}^{2}=\varsigma _{T}$, $\varsigma _{Q}\cdot \varsigma
_{E}=\varsigma _{R}$, $\varsigma _{R}\cdot \varsigma _{Q}=\varsigma
_{E}$, $\varsigma _{E}\cdot \varsigma _{R}=\varsigma _{Q}$.\ \ \ 

These then multiplied with the PCT transformations to define 6 additional
elements $\varsigma _{\alpha \beta }=\varsigma _{\alpha }\cdot \varsigma
_{\beta }$ with $\alpha \in \{P,T,C\}$ and $\beta \in \{Q,E,R\}$
labels of the abelian group. The transformations equations for these
follow immediately from (86) and (87). With\ \ these 13 elements,
the abelian discrete group closes.\ \ The multiplication table is
simply worked out using these multiplication rules\ \ 
\begin{equation}
\varsigma _{\alpha \beta }\cdot \varsigma _{\gamma \delta }= \varsigma
_{\alpha }\cdot \varsigma _{\beta }\cdot \varsigma _{\gamma }\cdot
\varsigma _{\delta }=\left( \varsigma _{\alpha }\cdot \varsigma
_{\gamma }\right) \cdot \left( \varsigma _{\beta }\cdot \varsigma
_{\delta }\right)  ,
\end{equation}

\noindent where $\alpha ,\gamma \in \{P,T,C\}$ and $\beta ,\delta
\in \{Q,E,R\}$.

The group elements $\varsigma _{P},\varsigma _{T},\varsigma _{Q},\varsigma
_{E}$ may be realized by the $4\times 4$ matrices.\ \ \ 
\begin{equation}
\begin{array}{ll}
 \varsigma _{P}\simeq \left( \begin{array}{llll}
 1 & 0 & 0 & 0 \\
 0 & -1 & 0 & 0 \\
 0 & 0 & -1 & 0 \\
 0 & 0 & 0 & 1
\end{array}\right) , & \varsigma _{T}\simeq \left( \begin{array}{llll}
 -1 & 0 & 0 & 0 \\
 0 & 1 & 0 & 0 \\
 0 & 0 & 1 & 0 \\
 0 & 0 & 0 & -1
\end{array}\right) , \\
 \varsigma _{Q}\simeq \left( \begin{array}{llll}
 1 & 0 & 0 & 0 \\
 0 & 0 & -1 & 0 \\
 0 & 1 & 0 & 0 \\
 0 & 0 & 0 & 1
\end{array}\right) , & \varsigma _{E}\simeq \left( \begin{array}{llll}
 0 & 0 & 0 & 1 \\
 0 & 1 & 0 & 0 \\
 0 & 0 & 1 & 0 \\
 -1 & 0 & 0 & 0
\end{array}\right) 
\end{array}
\end{equation}

\noindent with $\varsigma _{0}=I, \varsigma _{C}=-I $ and $\varsigma
_{R}=\zeta $. The remaining matrices may be computed from these
four elements using the group multiplication rules given above as
these four elements generate the group. These 4 elements generate
the group and so we need consider only these elements further. Elements
of the discrete group leave invariant the symplectic and Born-Green
orthogonal metrics and are automorphisms of $\mathcal{U}( 1,3) $group.
\begin{equation}
\begin{array}{l}
 \varsigma _{P}\cdot \Xi ( v,f,r,a) \cdot {\varsigma _{P}}^{-1}=\Lambda
( -v,-f,r,a) , \\
 \varsigma _{T}\cdot \Lambda ( v,f,r,a) \cdot {\varsigma _{T}}^{-1}=\Lambda
( -v,-f,r,a) , \\
 \varsigma _{C}\cdot \Lambda ( v,f,r,a) \cdot {\varsigma _{C}}^{-1}=\Lambda
( v,f,r,a) , \\
 \varsigma _{Q}\cdot \Lambda ( v,f,r,a) \cdot {\varsigma _{Q}}^{-1}=\Lambda
( -f,v,r,a) , \\
 \varsigma _{E}\cdot \Lambda ( v,f,r,a) \cdot {\varsigma _{E}}^{-1}=\Lambda
( -f,v,r,a) , \\
 \varsigma _{R}\cdot \Lambda ( v,f,r,a) \cdot {\varsigma _{R}}^{-1}=\Lambda
( -v,-f,r,a) 
\end{array}
\end{equation}

Finally, consider the finite discrete abelian group for extended
quaplectic case, we consider transformations of\ \ the basis $\{d
z, d \iota \}=\{\mathit{dt},\mathit{dq},\mathit{dp},\mathit{de}\}$
as embedded in the $6\times 6$ matrices. The above generators may
be embedded in the $6\times 6$ matrices and the additional generators
incorporating the additional discrete symmetry in (76) may be defined
as
\begin{equation}
\begin{array}{ll}
 \varsigma _{\alpha }^{+}= \left( \begin{array}{lll}
 \varsigma _{\alpha } & 0 & 0 \\
 0 & 1 & 0 \\
 0 & 0 & 1
\end{array}\right)  & \varsigma _{\alpha }^{-}= \left( \begin{array}{lll}
 \varsigma _{\alpha } & 0 & 0 \\
 0 & -1 & 0 \\
 0 & 0 & -1
\end{array}\right) 
\end{array}
\end{equation}

The full finite abelian discrete group $\mathcal{D}$ for the extended
quaplectic group is the 26 element abelian group with elements\ \ $\varsigma
_{\alpha }^{\pm }\in \mathcal{D}$,\ \ with $\alpha $ taking values
in the set of labels of the 13 elements of $\mathcal{D}\mbox{}^{\circ}$$
$. These transformations leave invariant the metric for the nonabelian
space defined by the Casimir invariant (82) and are automorphisms
of the quaplectic group. 

\section{Discussion}

\subsection{$n$ Dimensional Case and Limits}

For simplicity and clarity, we have studied the case with 1 position
dimension. The theory clearly generalizes to $n$ dimensions with\ \ \ 
\begin{equation}
\mathcal{C}( 1,n)  = \mathcal{U}( 1,n) \otimes _{s}\mathcal{H}(
n+1) = \text{}\text{}\left( \mathcal{U}( 1)  \otimes \mathcal{S}\mathcal{U}(
1,n) \right) \otimes _{s}\mathcal{H}( n+1) 
\end{equation}

\noindent where 
\begin{equation}
\left.  \mathcal{U}( 1,n) = \mathcal{O}( 2,2n) \right) \cap \mathcal{S}p(
2n+2) 
\end{equation}

\noindent In the physical case where $n=3$, $\mathcal{C}( 1,3) $
is $25$ dimensional.\ \ In the limiting case $b,c\rightarrow \infty
$,\ \ 
\[
\operatorname*{\lim }\limits_{b, c\,\rightarrow \:\infty } \mathcal{U}(
1,n)  = \mathcal{S}\mathcal{O}( n) \otimes _{s}\mathcal{H}( n) 
\]

\noindent Note that for $n=1$, 
\begin{equation}
\operatorname*{\lim }\limits_{b, c\,\rightarrow \:\infty }\mathcal{S}\mathcal{U}(
1,1)  = \mathcal{H}( 1) %
\label{contraction of su11 to h2}
\end{equation}

\noindent which is why (25) is the Heisenberg group composition
law for $\mathcal{H}( 1) $.\ \ This is the counterpart of the usual
relation for the orthogonal group
\begin{equation}
\operatorname*{\lim }\limits_{c\,\rightarrow \:\infty } \mathcal{S}\mathcal{O}(
1,n)  = \mathcal{E}( n) = \mathcal{S}\mathcal{O}( n) \otimes _{s}\mathcal{T}(
n) 
\end{equation}

\subsection{Comments on Quantum Mechanics}

A quantum theory may be constructed from the unitary representations
of a dynamical group. The wave equations and Hilbert space of basic
special relativistic quantum mechanics follow from the unitary irreducible
representations of the Poincar\'e group \cite{wigner1}. The unitary
representations of the Poincar\'e group determine infinite dimensional
Hilbert spaces that define the states of {\itshape free} particles.
The eigenvalue equations for the Hermitian representation of the
Casimir invariants are the basic field or wave equations of physics:
Maxwell, Dirac, Klein-Gordon and so forth.\ \ The eigenvalues labeling
the irreducible representations are the fundamental concepts of
mass and spin. 

One can likewise construct a quantum theory of the quaplectic group
by considering the unitary irreducible representations of the quaplectic
group.\ \ Again, the Hilbert space of particle states is determined
from these irreducible unitary representations. However, in this
case, these states include free and interacting particles with correspondingly
nonintertial frames. Again, the eigenvalue equations for the Hermitian
representation of the Casimir invariants are the basic field or
wave equations of physics of this theory and the eigenvalues must
define basic physical properties.\ \ 

This theory is presented in a companion paper \cite{low-Casimir
of Quaplectic}.\ \ The field equations are determined and shown
to satisfy certain basic criteria for being reasonable. The Schr\"odinger-Robinson
inqualities that generalize the\ \ Heisenberg uncertainty relations
may be shown to be invariant under the quaplectic group and this
may\ \ be used to study the semiclassical limit in terms of coherent
states \cite{jarvis morgan} The {\itshape scalar} case in this theory
is the relativistic oscillator.\ \ While the quaplectic invariant
wave equations have been determined in\ \ \cite{low-Casimir of Quaplectic},
these equations have not yet been explored to understand their physical
meaning. This is work that remains to be undertaken. 

\subsection{Summary}

There are many higher dimensional theories in the literature. The
theory presented is just a higher dimensional theory.\ \ Before
relativity, one would say that we live in a three dimensional space
with a universal time parameter. Relativity changed that to a four
dimensional {\itshape space-time}, or as we use the word space more
generally here, {\itshape position-time }manifold. Special relativity
eliminates the\ \ concept of an absolute rest frame but continues
to have the concept of an absolute inertial frame. Eliminating the\ \ absolute
rest frame requires the four dimensional space-time continuum that
no longer has an absolute sense of time.\ \ We simply generalized
this approach to noninertial frames. We consider the local transformations
on the time-position-momentum-energy space and show that the expected
transformations result under the nonrelativistic and special relativistic
assumptions. The general case of transformation between noninertial
frames results from requiring an invariant symplectic metric and
Born-Green orthogonal metric. An investigation of the equations
shows that the need for absolute inertial frame is eliminated, forces
and rates of change of energy are relative, and bounded by $b$ and
$b c$.\ \ \ For the physical case $n=3$,\ \ this group is $\mathcal{U}(
1,3) $.\ \ 

The basic wave equations of special relativistic quantum theory
arises from considering the unitary representations of the Poincar\'e
group (or more accurately, its universal cover).\ \ This leads us
to consider the inhomogeneous group on the\ \ time-position-momentum-energy
space.\ \ However,\ \ position and momentum and time and energy
do not commute. This\ \ leads us to consider the nonabelian Heisenberg
group that is the semidirect product of two translation groups.\ \ The
remarkable fact emerges that constructing a semidirect product with
the Heisenberg group as the normal group requires an invariant symplectic
metric. One need only hypothesize the Born-Green orthogonal metric.
The group $\mathcal{U}( 1,n) \otimes _{s}\mathcal{H}( 1+n) $ is
the quaplectic group.

The Born-Green orthogonal metric requires the introduction of a
new physical constant $b$ that can be taken to have the dimensions
of force. There are only three dimensionally independent physical
constants. These may be taken to be $c,b$ and $\hbar $.\ \ The remarkable
fact is that the quaplectic group introduces a relativity principal
that is reciprocal, in the sense of Born, to the usual special relativity.
Now, in addition to rates of change of position with time being
bounded by $c$, rates of change of momentum are bounded by $b$ and
rates of change of momentum are bounded by $b c$.\ \ The position-time
subspace is now observer frame dependent and these effects become
manifest for noninertial interacting particles with rates of change
of momentum approaching $b$. 

A quantum theory is constructed by considering the unitary representations.
The unitary irreducible representations determines the Hilbert space
of particle states. The eigenvalue equations of the Hermitian representations
of the Casimir invariant operators determines the field or wave
equations of the theory. The eigenvalues characterize basic particle
properties.\ \ This is discussed in a companion paper \cite{low-Casimir
of Quaplectic}. 

The nonabelian generalization of the time-position-momentum-energy
space that is characterized in this paper for the one dimensional
case has a rich and subtle structure. There is a very simple heuristic
model that may be helpful to visualize it.\ \ Special relativity
was greatly simplified with the introduction of the {\itshape four}
{\itshape vector} notation. One might think that the natural extension
here is, for the $n=3$ case, an {\itshape eight} or {\itshape ten}
vector\ \ notation.\ \ However, the nonabelian structure of the
space is best represented heuristically by a {\itshape quad. }Only
the degrees of freedom on each of the four faces of the quad commute.
Compare also with the quad of dimensions previously given in (51).
\[
\begin{array}{ll}
 T & Q \\
 P & E
\end{array}\ \ \ 
\]

The parity, time reversal, and the Born reciprocity transformations
that generate the discrete automorphism group have the action on
the quad given by 
\begin{gather*}
\varsigma _{P}:\begin{array}{ll}
 T & Q \\
 P & E
\end{array}\ \ \rightarrow \begin{array}{ll}
 \ \ \ \ T & -Q \\
 -P & \ \ \ \ E
\end{array} ,\ \ \ \ \ \ \varsigma _{T}:\begin{array}{ll}
 T & Q \\
 P & E
\end{array}\ \ \rightarrow \begin{array}{ll}
 -T & \ \ Q \\
 \ \ P & -E
\end{array}\ \ ,
\end{gather*}
\[
\varsigma _{Q}:\begin{array}{ll}
 T & Q \\
 P & E
\end{array}\ \ \rightarrow \begin{array}{ll}
 \ \ T & P \\
 -Q & E
\end{array},\ \ \ \ \ \ \ \ \ \varsigma _{E}:\begin{array}{ll}
 T & Q \\
 P & E
\end{array}\ \ \rightarrow \begin{array}{ll}
 -E & Q \\
 \ \ P & T
\end{array} 
\]

Now, continuing in this heuristic manner, we know that the universal
constant $c$ is associated with a relativity on the $(T,Q)$ and
$(P,E)$\ \ subspaces that is locally governed by the Lorentz group.
The remarkable fact is that the constant $b$ introduced above is
associated with a relativity on the\ \ $(T,P)$ and $(Q,E)$ subspaces
that is also locally governed by a Lorentz group. For this reason,
we call the relativity of velocity and forces {\itshape reciprocal
relativity}. Both of these are subgroups of the more general pseudo-unitary
group.\ \ 

The etymology of\ \ {\itshape quaplectic} is the following.\ \ \ \ {\itshape
Qua }is a seldom used English word that means {\itshape in the character
of} or simply {\itshape as} as in {\itshape Sin qua non}.\ \ {\itshape
Plectic} has origins in Greek which means {\itshape to pleat} or
{\itshape to fold diagonally}.\ \ So quaplectic means {\itshape
in the character of folding diagonally}.\ \ It also invokes {\itshape
Quad}, {\itshape Quantum} and {\itshape Symplectic}, all of which
play a role. 

The author thanks P. Jarvis for stimulating discussion and comments
on this paper and B. Hall for encouraging a simple physical description.

\section{Appendix}

\subsection{Comment on Lagrangian mechanics}\label{Section Lagrangian
mechanics}

Lagrangian mechanics introduces the {\itshape bias} to a position-time
manifold formalism.\ \ In this section, we show that the bias is
not intrinsic in the basic mathematical formulation as one can simply
Legendre transform also to an equivalent momentum-time Lagrangian
formulation.

We have been considering a formulation on $z =\{t,q,p,e\}\in \mathbb{P}\simeq
\mathbb{R}^{4}$ with frames $d z =\{d t,d q,d p, d e\}\in  {T^{*}}_{z}\mathbb{P}$.\ \ \ 

The symplectic 2-form $-\mathit{d}\tilde{e}\wedge \mathit{d}\tilde{t}+\mathit{d}\tilde{p}\wedge
\mathit{d}\tilde{q}$ in (23) may be integrated to define the 1-form
\begin{equation}
-e d \tilde{t}+ p d \tilde{q}=\left( -H( p,q,t) + p( t)  \frac{d
q( t) }{d t}\right) d t=L( q( t) ,\frac{d q( t) }{d t},t) 
\end{equation}

\noindent where Hamilton's equations $\frac{\mathit{dq}( t) }{\mathit{dt}}=\frac{\partial
H( q,p,t) }{\partial p}$ are used to solve for $p=p( q,\frac{\mathit{dq}}{\mathit{dt}})
$ provided that the Hessian is nonsingular\ \ (we use the general
expressions here which carry over to $n$ dimensions even though
in our current context we are in one dimension and the determinant
is trivial.)
\[
\mathrm{Det}\frac{\partial \mathit{H}( p,q,t\mathit{)} }{\partial
p \partial p}\neq 0
\]

\noindent The Lagrangian satisfies the variational principle $\delta
\int $$\mathit{L}\mathit{(} \mathit{q}\mathit{,}\frac{\mathit{dq}}{\mathit{dt}},t)
$=0 to yield the Euler-Lagrange equations
\begin{equation}
\frac{\partial L( q,v,t) }{\partial q}-\frac{\mathit{d}}{\mathit{dt}}\frac{\partial
L( q,v,t) }{\partial v}=0
\end{equation}

Geometrically, a number of things have happened here.\ \ Let $\mathbb{M}\simeq
\mathbb{R}^{2}$ with\ \ $(t,q)\in \mathbb{M}$\ \ and then $-e \mathit{dt}
+ p \mathit{dq}$ is an element of $T^{*}\mathbb{M} $.\ \ The Lagrangian
on the other hand, is a function on the tangent space $T \mathbb{M}$.\ \ This
is possible as $\{\mathit{dt}\}$ and $\{\mathit{dt},\mathit{dq}\}$
are invariant subspaces under the action of $\Phi $.\ \ That is,
there is an absolute notion of time that all observers agree on
and furthermore all observers agree on the position-time subspace
of the full space $\mathbb{P}$.\ \ 

Note also that $\{\mathit{dt},\mathit{dp}\}$ is an invariant subspace
and therefore all observers agree on the momentum-time subspace.
The corresponding integration of the symplectic two form is\ \ $-e
\mathit{dt} -q \mathit{dp}$ is an element of $T^{*}\check{\mathbb{M}}
$ with $\check{\mathbb{M}}\simeq \mathbb{R}^{2}$ with\ \ $(t,p)\in
\check{\mathbb{M}}$. Applying the transformations gives
\begin{equation}
-e d \tilde{t}- q d \tilde{p}=-\left( H( p,q,t) + q( t)  \frac{d
p( t) }{d t}\right) d t=L( p( t) ,\frac{d p( t) }{d t},t) 
\end{equation}

\noindent where Hamilton's equations $\frac{\mathit{dp}( t) }{\mathit{dt}}=-\frac{\partial
H( q,p,t) }{\partial q}$ are used to solve for $q=q( p,\frac{\mathit{dp}}{\mathit{dt}})
$ provided that the Hessian is nonsingular 
\[
\mathrm{Det} \frac{\partial \mathit{H}( p,q,t\mathit{)} }{\partial
q \partial q}\neq 0
\]

\noindent The Lagrangian satisfies the variational principle $\delta
\int $$\mathit{L}\mathit{(} \mathit{p}\mathit{,}\frac{\mathit{dp}}{\mathit{dt}},t)
$=0 to yield the Euler-Lagrange equations
\begin{equation}
\frac{\partial L( p,f,t) }{\partial p}-\frac{\mathit{d}}{\mathit{dt}}\frac{\partial
L( p,f,t) }{\partial f}=0
\end{equation}

Clearly, for a free particle, this latter Hessian is singular.\ \ However,
in the presence of long range $1/r^{2}$ type forces, there are no
truly free particles. Consequently, for classical systems with these
types of forces present, a free particle is an abstraction and in
principle one could formulate all of basic mechanics on this momentum-time
space. We are heavily biased by our Newtonian heritage to consider
{\itshape space}, that is {\itshape position space},{\itshape  }as
the fundamental arena of physics. This is re-enforced by special
relativity that brings time onto the same footing as position to
define space-time (that is {\itshape position-time space}).\ \ However,
the basic Hamilton's equations put momentum and position on completely
equal footing and this remains true in Dirac's nonrelativistic transformation
theory of quantum mechanics \cite{Dirac}. In fact, as just shown,
there is a reciprocal conjugate Lagrangian formulation on momentum-time
space that is equally valid in the nonrelativistic case as the position-time
Lagrangian formulation. 

In this heuristic sense, the simplest classical Hamiltonian formulation
satisfies Born reciprocity in position and momentum, there is no
distinction in the mathematical formulation that biases the formulation
to position-time over momentum-time.\ \ 

\subsection{Comment on the $ \mathcal{U}$(1,1) transformation equations}\label{Section
Hyperbolic u11}

The transformation equations $\mathcal{U}( 1,1) =\mathcal{U}( 1)
\otimes \mathcal{S}\mathcal{U}( 1,1) $ may also be written in terms
of hyperbolic trig functions.

The $\mathcal{S}\mathcal{U}( 1,1) $ transformation equations that
leave this invariant are 
\begin{equation}
\begin{array}{l}
 d\tilde{t}=\cosh  \omega  \mathit{dt} + \frac{\sinh  \omega }{\omega
}\left( \frac{\beta }{c}\mathit{dq}+ \frac{\gamma }{b}\mathit{dp}\mathrm{-}\frac{\vartheta
}{b c}\mathit{de}\right) , \\
 d\tilde{q}=\cosh  \omega  \mathit{dq} + \frac{\sinh  \omega }{\omega
}\left( c \beta \mathit{dt}-\frac{\gamma }{b} \mathit{de} \mathrm{+}\frac{c
\vartheta }{b}\mathit{dp}\right) , \\
 d\tilde{p}=\cosh  \omega  \mathit{dp} + \frac{\sinh  \omega }{\omega
}\left( b \gamma \mathit{dt}+\frac{\beta }{c}\mathit{de} \mathrm{-}\frac{b
\vartheta }{c}\mathit{dq}\right)  \\
 d\tilde{e}=\cosh  \omega  \mathit{de} + \frac{\sinh  \omega }{\omega
}\left( -b \gamma \mathit{dq}+c \beta  \mathit{dp}\mathrm{+} \mathit{b}
\mathit{c} \vartheta  \mathit{dt}\mathit{\ \ }\right) ,
\end{array}
\end{equation}

\noindent where\ \ $\omega ^{2}=\beta ^{2}+\gamma ^{2}-\vartheta
^{2}$.\ \ Note immediately that if $\gamma =\vartheta =0$, these
reduce to the usual special relativity equations. It may be directly
verified that these equations leave invariant the symplectic metric
\begin{equation}
\begin{array}{rl}
 -d\tilde{e}\wedge \mathit{d}\tilde{t}\mathrm{+}\mathit{d}\tilde{p}\wedge
\mathit{d}\tilde{q} & =-\mathit{de}\wedge \mathit{dt}\mathrm{+}\mathit{dp}\wedge
\mathit{dq}
\end{array}
\end{equation}

\noindent and the Born-Green metric line element in (30). The group
is a matrix group with elements\ \ $d\tilde{z} =\Xi ( \beta ,\gamma
,\vartheta ) \mathit{dz}$\ \ realized by the matrix
\begin{equation}
\Xi ( \beta ,\gamma ,\vartheta ) =\left( \begin{array}{llll}
 \cosh  \omega   & \frac{\beta  \sinh  \omega }{\omega  c} & \frac{\gamma
\sinh  \omega }{\omega  b} & -\frac{\vartheta  \sinh  \omega }{\omega
b c} \\
 \frac{c \beta  \sinh  \omega }{\omega  } & \cosh  \omega   & \frac{c\ \ \vartheta
\sinh  \omega }{b \omega  } & \frac{-\gamma  \sinh  \omega }{\omega
b} \\
 \frac{b \gamma  \sinh  \omega }{\omega  } & -\frac{b \vartheta
\sinh  \omega }{c \omega } & \cosh  \omega   & \frac{\beta  \sinh
\omega }{c \omega  } \\
 \frac{b c \vartheta  \sinh  \omega }{\omega  } & \frac{-b \gamma
\sinh  \omega }{\omega  } &  \frac{c \beta  \sinh  \omega }{\omega
}  & \cosh  \omega  
\end{array}\right) %
\label{Quaplectic matrix}
\end{equation}

\noindent In order that the transformations are hyperbolic and not
the usual sine or cosine functions that would result in oscillations,
we must have\ \ \ $\omega ^{2}=\beta ^{2}+\gamma ^{2}-\vartheta
^{2}\geq 0$.\ \ \ This formalism is equivalent to the discussion
in the\ \ main body of text by defining the parameterization
\begin{equation}
v=\frac{c \beta }{\omega } \tanh  \omega ,\ \ f=\frac{b \gamma }{\omega
} \tanh  \omega  , r=\frac{b c \vartheta  }{\omega } \tanh  \omega
\end{equation}

\noindent Then, define
\begin{equation}
w^{2}=\ \ {\left( \frac{v}{c}\right) }^{2}+ {\left( \frac{f}{c}\right)
}^{2}- {\left( \frac{r}{b c}\right) }^{2}= {\left( \tanh  \omega
\right) }^{2}
\end{equation}
Note that $\omega \geq 0$ implies that $d s^{2}\leq 0$. It follows
immediately that $\cosh  \omega  ={(1-w^{2})}^{-1/2}$ and\ \ $\sinh
\omega  =w {(1-w^{2})}^{-1/2}$ . Also note that 
\begin{equation}
\frac{ \beta }{\omega }= \frac{v}{c w},\ \ \frac{ \gamma }{\omega
}= \frac{f}{b w},\ \ \frac{\ \ \vartheta }{\omega }= \frac{r}{b
c w}
\end{equation}

\noindent Equation (32) follows directly.

\end{document}